\documentclass[%
aps,
prx,
reprint,
superscriptaddress,
nofootinbib,
amsmath,
amssymb
]{revtex4-2}
\usepackage{
    physics,
    graphicx,
    multirow,
    tabularx,
    mathtools, 
    placeins, 
    nicefrac, 
    hyperref, 
    threeparttable, 
    siunitx,
    enumitem,
    makecell,
    stmaryrd,
    booktabs
}

\usepackage[dvipsnames]{xcolor}
\hypersetup{
    colorlinks,
    linkcolor={blue!50!black},
    citecolor={blue!50!black},
    urlcolor={blue!80!black}
}

\usepackage[caption=false]{subfig} 
\usepackage[capitalize]{cleveref}
\Crefname{section}{Sec.}{Secs.}
\crefname{appendix}{Appendix}{Appendices}
\Crefname{appendix}{Appendix}{Appendices}
\crefname{subappendix}{Appendix}{Appendices}
\Crefname{subappendix}{Appendix}{Appendices}

\newcommand{\red}[1]{{\color{red} #1}}

\begin{document}

\title{NeST: ultra-high-rate qLDPC codes with sub-$50\mu$s stabiliser cycles via spin shuttling}

\newcommand{\qmaddress}{\affiliation{Quantum Motion, 9 Sterling Way, London N7 9HJ, United Kingdom}}
\newcommand{\oxddress}{\affiliation{Department of Materials, University of Oxford, Parks Road, Oxford OX1 3PH, United Kingdom}}
\newcommand{\uclddress}{\affiliation{Department of Physics and Astronomy, University College London, Gower St, London WC1E 6BT, United Kingdom}}

\newcommand{\comHamza}[1]{{\color{brown} #1}}

\author{Adam Siegel}
\email{adam@quantummotion.tech}
\qmaddress


\date{\today}

\begin{abstract}
Quantum low-density parity-check codes now reach encoding rates up to $50\%$ at useful distances. Their checks however are long-range group actions, not local neighbourhoods, and that connectivity has so far been supplied by neutral atoms. Silicon spin qubits offer the same primitive --- coherent transport of qubit blocks --- through electron shuttling, and on microsecond rather than millisecond timescales. They however travel along fabricated tracks rather than free space: permutations an atom array obtains by moving tweezers must here become collision-free motions on a physical lattice. We here present the first code and spin-qubit hardware co-design. We build an architecture whose geometry mimics that of a family of non-abelian groups, broad enough to host codes of excellent parameters: a hierarchy of nested conveyor-belt loops on a plain square lattice of shuttling tracks, in which every group multiplication becomes a rigid, globally controlled motion of whole qubit trains. We call it NeST, for Nested Shuttling Tracks. A code search over the family returns rate-$20\%$ mitten codes up to $\llbracket 1400,280,\leq 28\rrbracket$, and circuit-level simulations including shuttling dephasing give thresholds above $10^{-3}$ with sub-$50\mu$s stabiliser cycles, two to three orders of magnitude faster than neutral-atom implementations. Via extrapolations, we estimate that this largest code may reach a logical error probability per qubit per round of $10^{-14}$, assuming a ten-fold improvement on early shuttling demonstrations, with other physical errors at $p=0.1\%$. Logic is compatible with NeST too: for the $\llbracket 150,30,10\rrbracket$ code we build a code-surgery gadget in which every qubit again moves only by rigid belt motions, at no qubit cost and a $29\%$ increase in gate layers. This demonstrates that ultra-low-overhead fault tolerance and a fast clock speed are not competing goals, and that silicon spin qubits can supply both.
\end{abstract}
\maketitle

\section{Introduction}

\begin{figure*}
    \centering
    \includegraphics[width=\linewidth]{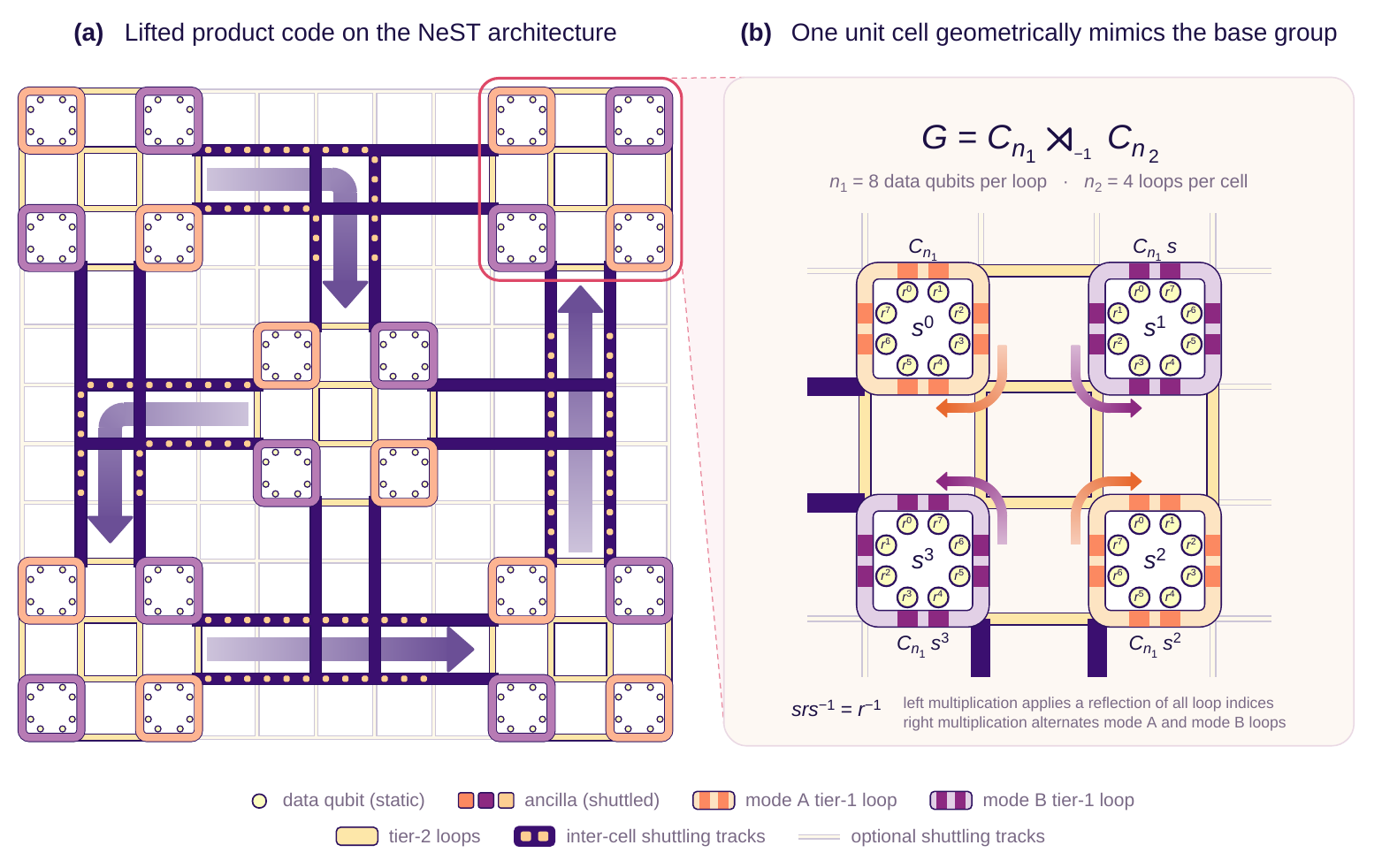}
    \caption{\textbf{A lifted product code on the NeST architecture.} \textbf{(a)} The five data qubit blocks of a mitten code --- four left blocks (corners) and one right block (centre) --- become five \textit{unit cells} on a single square lattice of shuttling tracks, arranged in the 2D pattern of the protograph. Each cell holds $|G|$ static data qubits and can host one block of shuttled ancillas at a time. Within a cell, ancillas move as described in \cref{fig:motions}; between cells they are translated wholesale along the inter-cell highways (dark purple, arrows), which supply the long-range connectivity the base matrices $A$ and $B$ may demand. The pale tracks are optional: the lattice is strictly square throughout and may be made denser to shorten routes or avoid defects, or sparser to leave room for control electronics. \textbf{(b)} Zoom on one unit cell: $n_2=4$ tier-1 conveyor loops of $n_1=8$ data qubits (yellow circles) each, with one ancilla site (squares) beside every data qubit. Qubits are labelled with the elements of $G=C_{n_1}\rtimes_{-1}C_{n_2}$ (\cref{eq:2_tier_group}), with each loop carrying the subgroup $C_{n_1}s^l=\{r^ks^l\}_{0\leq k<n_1}$. The geometry mirrors the group law: a multiplication by $r$ or $s$ respectively advances tier-1 or tier-2 loops by one increment. Right multiplications alternate the relative rotation direction of mode A and B loops, while left multiplications reflect the loop indices, $r^k\mapsto r^{-k}$, which the hardware realises by letting the ancilla train enter its destination loop in the opposite direction (panels (d) and (e) of \cref{fig:motions}).}
    \label{fig:main_figure}
\end{figure*}

Quantum error correction buys reliability with redundancy, and the exchange rate is set by the code. The surface code has long been the default choice for solid-state platforms because it asks only for 2D nearest-neighbour interactions, but it encodes a single logical qubit in $O(d^2)$ physical ones~\cite{Fowler_sc}, so the physical qubit count dominates the cost of any large computation. More general quantum low-density parity-check (qLDPC) codes break that scaling: by giving up geometric locality they encode a constant fraction of their physical qubits, often at the cost of higher stabiliser weights. Hypergraph~\cite{tillich_hypergraph_2014}, lifted~\cite{panteleev_kalachev_lifted_2021} and balanced~\cite{breuckmann_balanced_2021} products provide the general construction, and explicit instances have moved quickly from asymptotic statements to useful block sizes. For instance, bivariate-bicycle codes reach a rate of $1/12$ at distance 12~\cite{bravyi_bb_2024}, while recent lifted products over non-abelian groups achieve rates of $20$--$30\%$ at distances in the twenties~\cite{bhardwaj2026highrateqldpcprocessors, cain2026shorsalgorithmpossible10000}. Codes based on affine permutations even managed a rate higher than $50\%$~\cite{lu2026quantumerrorcorrectionultralow}. At those values a logical qubit costs order ten or less physical qubits rather than a thousand.

What these ultra-high-rate constructions ask for in return is connectivity of a very particular kind. Their checks are not local neighbourhoods but long-range \textit{group actions}: each one couples a block of ancillas to a block of data qubits permuted by a single group element. A platform that can physically transport whole blocks of qubits therefore implements them almost for free, and this is precisely what reconfigurable neutral-atom arrays do by moving atoms with optical tweezers~\cite{bluvstein_coherent_transport_2022, bluvstein_logical_2024}. Unsurprisingly, the codes and the hardware have co-evolved there, and the highest-rate constructions published to date were designed with atom arrays in mind~\cite{bhardwaj2026highrateqldpcprocessors, cain2026shorsalgorithmpossible10000, lu2026quantumerrorcorrectionultralow}. The premise of this paper is that the enabling primitive --- moving qubits in bulk, coherently --- is not exclusive to atoms. Silicon spin qubits have it too, in the form of electron shuttling. The analogy stops there, however, and the gap it leaves is the subject of this work. An atom array transports its qubits through free space: any permutation is a matter of where the tweezers are told to go, the qubits pass over one another without consequence, and the connectivity graph is limited only by the control optics. Electrons are not so free. They move along fabricated tracks, so the set of realisable permutations is fixed at lithography time rather than chosen at runtime, and two electrons meeting on the same track is a forbidden collision. Possessing coherent block transport is therefore necessary but not sufficient: the group actions an atom array performs by moving tweezers must here be decomposed into motions that a physical lattice can route, synchronously and without collisions. Whether that is possible at all, for a group rich enough to support a high-rate code, is not obvious a priori.

Spin qubits are in other respects a mature platform. They offer long coherence times~\cite{Struck_2020} with a qubit footprint of order $100$~nm~\cite{gonzalez-zalbaScalingSiliconbasedQuantum2021}, are fabricated with the same processes as classical microelectronics~\cite{Maurand_2016, Zwerver_2022, steinacker2024300mmfoundrysilicon, swift_large_scale_2025, swift_superinductor_2025, clarke_readout_2025} and sit comfortably beside cryogenic control circuitry~\cite{XueXiao2021Ccco, ruffino2021integrated, thomas_rapid_2025}. Compact multi-qubit processors have been operated~\cite{Philips_2022} and small error-correcting codes demonstrated~\cite{Takeda_2022}, with digitally controlled multi-qubit processing units now reported at larger scale~\cite{hrl2026qpu, universalFourSingletTriplet, thomas_rapid_2025}, with one- and two-qubit gate fidelities above the surface-code threshold~\cite{Xue_2022, Noiri_2022, Mills_2022} and in places beyond $99.9\%$~\cite{yonedaQuantumdotSpinQubit2018}. CZ and CNOT gates are both native, offering multiple avenues for stabiliser circuits. Shuttling is the newer capability and the decisive one here: electrons have been moved coherently across a device at $99.99\%$ fidelity per increment and at high speed~\cite{desmet2024highfidelitysinglespinshuttlingsilicon, binder_shuttling_2025}, with theory suggesting there is further headroom~\cite{Langrock_2023, binder_shuttling_2025}. Beyond extending interaction ranges, motion has been shown to average out dephasing~\cite{burkardSemiconductorSpinQubits2021} and to align qubit frequencies for global control~\cite{jnane2025harnessingelectronmotionglobal}. Junctions have also been demonstrated, via the conveyor-mode routing of single electrons between two independently driven shuttle lanes, with an inter-lane charge transfer infidelity below $10^{-6}$ and, notably, no control lines beyond the four channels per belt~\cite{beer2026tjunction}. 

A substantial body of architecture work already exploits this capability, replacing the static quantum computer with one in which qubits are permanently in motion. The Crossbar~\cite{li_crossbar_2018}, Spiderweb~\cite{boterSpiderwebArraySparse2022} and SpinBus~\cite{kunne_spinbus_2024} proposals shuttle data and ancilla qubits between manipulation zones to extend interaction ranges and free up area for control electronics~\cite{pataki_surface_code_crossbar_2024, escofet2025quantumreversemappingsynthesizing}; the Looped Pipeline~\cite{cai_looped_pipelines_2023} circulates both species around short loops to fold a 3D stack of codes into a 2D lattice; and the $2\times N$~\cite{siegel_two_by_n_2024, micciche2025optimizingcompilationerrorcorrection} and Snakes on a Plane~\cite{siegel2025snakesplanemobilelow} architectures reformat logical qubits into 1D filaments. Almost all of them, however, aim at the surface code. Ref.~\cite{chadwick2026cablecarefficientlyschedulingqldpc}'s CAbLECAR is an example treating high-rate qLDPC codes, but the approach remained deliberately code-agnostic: they schedule an arbitrary check matrix onto a fixed lattice and pay for that generality in routing depth. Spin qubits have therefore not yet been given an ultra-high-rate code of their own.

In this work we close that gap by co-designing the code and the hardware. Rather than asking how to route an existing code's permutations through a track network, we ask which groups a track network can route for free, and build the code from those. Our starting point is what a looped conveyor belt does naturally: it advances every qubit it carries by the same number of sites, in either direction, under a single global control signal and regardless of how many qubits are on it. This is a cyclic shift, meaning that one looped track can implement multiplications in a cyclic group cheaply. Arranging several such loops in a ring and passing qubits from one to the next repeats the construction one level up, and nesting it further builds a product of $p$ cyclic groups --- $p$ levels of rotations. Crucially, we show that \textit{non-abelianness} (beneficial to bypass caps on the code distance~\cite{smarandache_vontobel_2012}) can be added at no extra cost in the form of the inversion $r \mapsto r^{-1}$, which is simply obtained by choosing in which direction a group of ancilla enters its destination loop (panels (d) and (e) of \cref{fig:motions}). We deduce a large family of groups $G(\vec{n},\sigma)$ (\cref{eq:G_n_sigma}) for which left and right group actions reduce to a rigid motion of whole belts: every qubit moves synchronously without the need for individual addressability, and nothing is permuted in place. We call the resulting layout NeST, for Nested Shuttling Tracks; it is shown in \cref{fig:main_figure}.

We stress that NeST requires nothing of the hardware that proposals cited above do not already ask. The substrate is a plain lattice of linear shuttling tracks interfaced at junctions --- where four-way junctions may be replaced with pairs of three-way junctions if needed. Half the qubits (ancillas) are shuttled through the tracks while the other half (data) are kept static along them. The nested loops are a functional mode of that lattice not an extra requirement, and the only primitive used is conveyor-mode shuttling of the kind already demonstrated~\cite{Seidler_2022, xue_qubus_2024, desmet2024highfidelitysinglespinshuttlingsilicon}.

Our results are fourfold. First, on the theoretical side, the group family $G(\vec{n},\sigma)$ contains every finite abelian group, so every quasi-cyclic lifted-product code is implementable on NeST --- including the $30\%$-rate, distance-24 constructions of Ref.~\cite{cain2026shorsalgorithmpossible10000}. Along with them are non-abelian members that, at equal check weight, are not bound by the distance cap of Ref.~\cite{smarandache_vontobel_2012}. Second, we look for novel non-abelian codes and run a deep code search over mitten codes~\cite{bhardwaj2026highrateqldpcprocessors} inside our implementable group family. We report rate-$20\%$ candidates up to $\llbracket 1400,280,\leq 28 \rrbracket$ (\cref{tab:target_codes}). Third, we simulate those codes end to end on the architecture --- gate ordering optimised for high circuit distance and low shuttling, every ancilla motion routed and costed, shuttling dephasing included --- and find thresholds above $10^{-3}$ together with syndrome-extraction cycles of a few tens of microseconds. Extrapolating to a $0.1\%$ machine and assuming a ten-fold improvement of shuttling fidelities over early demonstrations, we estimate NeST to enable logical error rates as low as $10^{-14}$ with a distance-27 mitten code, requiring only $9$ physical qubits per logical qubit, against $1457$ for the surface code. Fourth, we address logic. Measuring logical operators on a qLDPC code requires attaching a surgery gadget, which deforms the code into one that may not be NeST-compatible without further work. We show that this can be avoided: for the $\llbracket 150,30,10\rrbracket$ code we construct a gadget whose every qubit still moves by rigid, globally controlled belt motions, using exactly as many qubits as the unconstrained gadget of Ref.~\cite{bhardwaj2026highrateqldpcprocessors} and preserving the code distance.

The paper is organised as follows. \cref{sec:lifted_product} fixes notation for lifted product codes. \cref{sec:architecture} builds NeST, from a single cyclic belt up to the $p$-tier layout. \cref{sec:perf_eval} selects codes, compiles their stabiliser circuits, maps them onto the lattice and reports logical error rates. \cref{sec:logic} turns to logic and builds a NeST-compatible surgery gadget. We close with the outlook of \cref{sec:discussion}.

\section{Lifted product codes} \label{sec:lifted_product}

In this section, we briefly review the construction of lifted product codes and introduce relevant parameters. This may be skipped by knowledgeable readers.

Let $G$ be a finite group of order $|G|$ and let $\mathbb{F}_2[G]$ be its group algebra over the binary field, \textit{i.e.} the set of formal sums $a=\sum_{g\in G}a_g\,g$ with $a_g\in\mathbb{F}_2$, added coefficient-wise and multiplied by extending the group law linearly. Each such $a$ gives rise to two $|G|\times|G|$ binary matrices, those of left and right multiplication by $a$ in the basis $\{g\}_{g\in G}$,
\begin{equation} \label{eq:regular_reps}
    \big[L(a)\big]_{x,y}=a_{xy^{-1}}, \qquad \big[R(a)\big]_{x,y}=a_{y^{-1}x},
\end{equation}
so that $L(a)$ and $R(a)$ are sums of $|a|\equiv\sum_g a_g$ permutation matrices, $|a|$ being the \textit{weight} of $a$. The single property we shall need is that the two actions commute,
\begin{equation} \label{eq:LR_commute}
    L(a)\,R(b) = R(b)\,L(a) \qquad \forall\, a,b\in\mathbb{F}_2[G],
\end{equation}
which is nothing but the associativity of the group law. It is also convenient to introduce the antipode $a^{*}=\sum_g a_g\,g^{-1}$, extended entry-wise and together with a transpose to matrices over $\mathbb{F}_2[G]$ as $A^{\dagger}\equiv(A^{*})^{\top}$. One can check from \cref{eq:regular_reps} that $L(a)^{\top}=L(a^{*})$ and $R(a)^{\top}=R(a^{*})$.

A lifted product code~\cite{panteleev_kalachev_lifted_2021} is specified by two \textit{base matrices} over the group algebra, $A\in\mathbb{F}_2[G]^{m_A\times n_A}$ and $B\in\mathbb{F}_2[G]^{m_B\times n_B}$, and is the CSS code with parity check matrices
\begin{equation} \label{eq:parity_check_matrix}
\begin{aligned}
    H_X &= \big[\;L(A)\otimes I_{n_B} \;\big|\; I_{m_A}\otimes R(B^{\dagger})\;\big],\\[2pt]
    H_Z &= \big[\;I_{n_A}\otimes R(B) \;\big|\; L(A^{\dagger})\otimes I_{m_B}\;\big],
\end{aligned}
\end{equation}
where $L(A)$ denotes the $m_A|G|\times n_A|G|$ matrix obtained by replacing every entry $a_{ij}$ of $A$ with the block $L(a_{ij})$, and likewise for $R(B)$. The CSS condition follows immediately: using $L(A^{\dagger})^{\top}=L(A)$, $R(B)^{\top}=R(B^{\dagger})$ and \cref{eq:LR_commute}, the two terms of $H_XH_Z^{\top}$ are both equal to $L(A)\otimes R(B^{\dagger})$ and cancel over $\mathbb{F}_2$. Lifting the left sector with $L$ and the right sector with $R$ is what makes the construction work for non-abelian $G$; for abelian $G$ the two coincide and one recovers the quasi-cyclic lifted products of Refs.~\cite{panteleev_kalachev_lifted_2021, bravyi_bb_2024}, and for $m_A=n_A=m_B=n_B=1$ the two-block group algebra codes of Ref.~\cite{2BGA_codes}.

Reading off the block structure of \cref{eq:parity_check_matrix}, the code has
\begin{equation} \label{eq:lp_params}
    n = (n_An_B + m_Am_B)\,|G|
\end{equation}
physical qubits, split into a \textit{left} sector of $n_An_B$ blocks and a \textit{right} sector of $m_Am_B$ blocks of $|G|$ qubits each, together with $m_An_B$ blocks of $X$ checks and $n_Am_B$ blocks of $Z$ checks. Every check has weight $w=\max_i\sum_j|a_{ij}| + \max_i\sum_j|b_{ij}|$ at most, so the code is LDPC as long as the base matrices are sparse and $G$ is grown at fixed $(A,B)$ shape and weight. The number of logical qubits $k$ is fixed by the ranks of the lifted blocks and is in general computed numerically.

Two features of \cref{eq:parity_check_matrix} drive everything that follows. First, each block of $H_X$ or $H_Z$ is a sum of at most $\max_{ij}|a_{ij}|$ (or $|b_{ij}|$) permutation matrices, each of which permutes a block of $|G|$ ancillas against a block of $|G|$ data qubits by a \textit{single} group element, acting from the left in the left sector and from the right in the right sector. Implementing a code block therefore reduces entirely to implementing left and right multiplications by group elements --- which is the problem \cref{sec:architecture} solves. Second, the block structure is inherited by the hardware: the $n_An_B+m_Am_B$ qubit blocks become the unit cells of \cref{fig:main_figure}, laid out in the 2D pattern of the \textit{protograph}.

For the numerical simulations, this paper focuses on the mitten codes of Ref.~\cite{bhardwaj2026highrateqldpcprocessors}, which take $A$ and $B$ both of shape $1\times 2$ with weight-3 entries, $A=\begin{pmatrix}a_0 & a_1\end{pmatrix}$ and $B=\begin{pmatrix}b_0 & b_1\end{pmatrix}$. \cref{eq:lp_params} then gives $n = 5|G|$ qubits in five blocks --- four left, one right --- with two blocks each of $X$ and $Z$ checks of weight $w=9$, and $k=|G|$ whenever one entry of each base matrix is chosen to lift to a full-rank block, so that the rate is $k/n=1/5$ by construction. Writing out \cref{eq:parity_check_matrix} in this case, and labelling the four left data blocks $D_1,\ldots,D_4$, the right one $D_5$, and the check blocks $X_0$, $X_1$, $Z_0$ and $Z_1$,
\begin{equation} \label{eq:parity_check_mitten}
\begin{aligned}
    H_X &= \bordermatrix{ & D_1 & D_2 & D_3 & D_4 & D_5 \cr
        X_0 & L(a_0) & 0 & L(a_1) & 0 & R(b_0^{*}) \cr
        X_1 & 0 & L(a_0) & 0 & L(a_1) & R(b_1^{*})}, \\[10pt]
    H_Z &= \bordermatrix{ & D_1 & D_2 & D_3 & D_4 & D_5 \cr
        Z_0 & R(b_0) & R(b_1) & 0 & 0 & L(a_0^{*}) \cr
        Z_1 & 0 & 0 & R(b_0) & R(b_1) & L(a_1^{*})}.
\end{aligned}
\end{equation}
Each entry is a $|G|\times|G|$ block, and each is a sum of three permutation matrices.

\section{The NeST architecture} \label{sec:architecture}

As we have just noted, lifted products are based on group actions. As these can be understood as permutations of the group elements, neutral atoms are a straightforward candidate for their implementation, since they can conveniently swap rows and columns of qubits with little cost. While such an operation is not directly available in spin qubit devices, these however support electron shuttling. Multiple architectures capitalising on such a capability have been proposed \cite{li_crossbar_2018, boterSpiderwebArraySparse2022, pataki_surface_code_crossbar_2024, kunne_spinbus_2024, escofet2025quantumreversemappingsynthesizing, cai_looped_pipelines_2023, siegel_two_by_n_2024, micciche2025optimizingcompilationerrorcorrection, siegel2025snakesplanemobilelow}, always taking the form of a latticework of linear shuttling tracks connected at junctions. In this paper and in particular in this section, we show that with the proper interpretation of the tracks of a simple square lattice, one can mimic the geometric representation of specific mathematical groups. This will give rise to the NeST architecture and ultimately, will allow us to embed wide classes of lifted product codes, including high-performance mitten codes \cite{bhardwaj2026highrateqldpcprocessors}. Note that if preferable, the four-way junctions a square lattice requires may simply be replaced with two consecutive three-way junctions, which have recently been demonstrated experimentally~\cite{beer2026tjunction}.

To assess the quality of our embedding, we will require the following guidelines:
\begin{itemize}
    \item \textit{Moderate  shuttling distance}: as shuttling can incur errors and is not instantaneous, this provides fast and low-noise stabiliser cycles. We allocate a maximum per-qubit shuttling budget $n\,w\,\ell_{qq}$, where $n$ is the number of physical qubits of the code, $w$ the stabilisers weight and $\ell_{qq}$ the inter-qubit distance.
    \item \textit{Globally controlled movement}: shuttling each qubit with individual controls represents a formidable experimental challenge: where possible subgroups of electrons should instead be shuttled collectively in a conveyor-belt mode \cite{Seidler_2022, xue_qubus_2024}.
    \item \textit{High synchronisation}: qubits should reach their destinations at mostly the same time, and ancilla should preferably not be left idling while others are involved in a two-qubit gate. This facilitates low idling errors and faster stabiliser cycles.
\end{itemize}

In all the following, we will assume that spins can be shuttled at a speed $v=50$ m/s. Each edge of the square grid is taken to be $\ell_e=1~\mu$m long to accommodate for the junctions' electrodes~\cite{beer2026tjunction}. Since a tier-1 loop is one grid square carrying $n_1$ data qubits (except in \cref{sec:logic}), the inter-qubit distance follows as $\ell_{qq}=4\ell_e/n_1$, between $0.6$ and $1.3~\mu$m for the codes considered here. The tracks are therefore sparser than a dense dot array would be, which costs cycle time and shuttling distance, but buys the room that junctions, readout regions and control electronics all require. Note that if fabrication rules led to the use longer square grid edges, then although cycles would extend the opportunity would arise for loading more qubits in each loop. Finally, data qubits will always be kept static, whereas ancillas are shuttled. This design choice is relatively arbitrary but will help reduce the impact of shuttling dephasing noise (with CZ-based stabiliser circuits it cannot spread to data qubits).

Throughout a stabiliser cycle, the interactions an ancilla qubit participates in are encoded in the parity check matrix (see \cref{eq:parity_check_matrix}). Each block matrix is a sum of permutations representing a left or right group multiplication. Therefore, the implementation of a given block requires the capability to permute the ancilla qubits with respect to the data qubits. Although generic permutations could theoretically be generated via qubit shuttling, this might incur significant overheads, in terms of shuttling time and noise. Rather, we here demonstrate that specific groups can naturally be mapped to shuttling tracks, respecting all the above guidelines.

\subsection{Embedding the cyclic group}

\begin{figure}
    \centering
    \includegraphics[width=\linewidth]{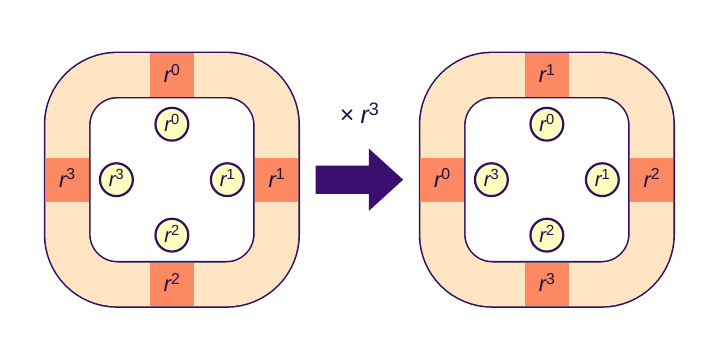}
    \caption{\textbf{Cyclic group embedding on a conveyor loop.} A single loop with $n = 4$ sites, drawn before (left) and after (right) the multiplication by $r^3$. The data qubits (circles) are static and label the sites of the loop by the group elements $r^0,\dots,r^{3}$; the ancillas (squares) ride the belt and are labelled by the element they carry. Advancing the belt by three sites sends the ancilla $r^j$ from site $r^j$ to site $r^{j+3}$, leaving every data qubit where it is. Thus a multiplication by $r^k$ is realised by one global rotation of the belt, meaning that the shuttling control cost is independent of $n$.}
    \label{fig:cyclic_group}
\end{figure}

The simplest group one can implement is the cyclic group,
\begin{equation}
    C_n = \{r^i,\, i\in \llbracket0,n-1\rrbracket\}
\end{equation}
where $n$ is a positive integer and $r$ is the group generator with $r^n=1$. It is implemented in \cref{fig:cyclic_group}, where a circular shuttling track hosts ancilla qubits, and static data qubits are placed inside. Every data and ancilla qubit is labelled by a group element $g=r^k$, such that the indices are aligned in the initial configuration (left panel). Applying a given group action, say a multiplication by $r^3$, means reshuffling the ancillas such that ancilla $r^k$ is now adjacent to data $r^{k+3}$. Straightforwardly, this is implemented by rotating the ancilla by three increments in the clockwise direction, or more directly, one increment in the anti-clockwise direction (right panel). In this example, ancillas are all moved collectively. Besides, the alignment of the ancilla and data indices guarantees the parallelisability of the two-qubit gates and no latency, as per the guidelines we laid out.

\subsection{Semidirect product of two cyclic groups}

\subsubsection{Shuttling tracks layout}

Abelian codes built on cyclic groups can be excellent --- Ref.~\cite{cain2026shorsalgorithmpossible10000} reaches a $30\%$ rate at distance 24 --- but they get there through large protographs and high check weight, which an architecture pays for in gates and in shuttling. Non-abelian groups offer a second route to the same place~\cite{smarandache_vontobel_2012, 2BGA_codes, bhardwaj2026highrateqldpcprocessors}: distance from the structure of the group rather than from the size of the base matrix. This motivates the study of more complex groups. One key observation is that if the groups $G$ and $G'$ are efficiently implementable (\textit{i.e.} follow our three guidelines), then their product should be too. In practice, the system would simply look like a collection of interconnected loops like the one represented in \cref{fig:cyclic_group}. Beyond simple direct products that preserve the abelianness, we demonstrate that the following semidirect product of cyclic groups is efficiently implementable:
\begin{equation} \label{eq:2_tier_group}
    C_{n_1} \rtimes_{-1} C_{n_2} = \{r^is^j,\, i\in \llbracket0,n_1-1\rrbracket,\, j\in \llbracket0,n_2-1\rrbracket\}
\end{equation}
where $r$ and $s$ are the group generators with:
\begin{align}
    r^{n_1}&=1\\
    s^{n_2}&=1\\
    srs^{-1}&=r^{-1}
\end{align}
Importantly, this last relation, which may be rewritten as $sr=r^{-1}s$, defines the commutation rule between $r$ and $s$. $n_1$ can be any integer $n_1\geq3$ but $n_2$ is restricted to even values. This is because conjugation by $s$ acts on the rotations as the inversion $r\mapsto r^{-1}$, an automorphism of order exactly $2$ for $n_1\geq3$. Conjugating $n_2$ times must return every element to itself, forcing $n_2$ to be even.

\begin{figure}
    \centering
    \includegraphics[width=\linewidth]{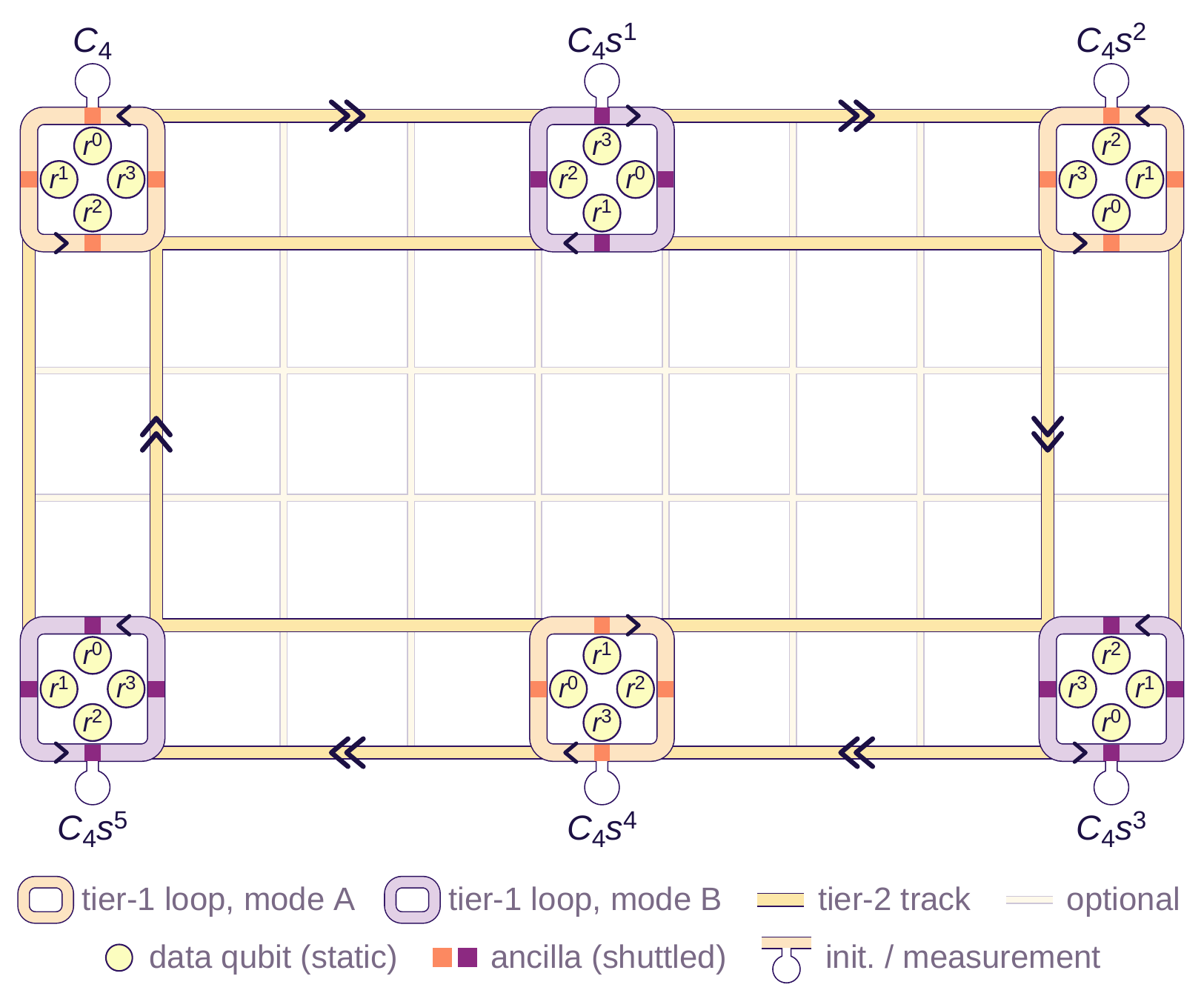}
    \caption{\textbf{Shuttling-track embedding of the group $G = C_{n_1} \rtimes_{-1} C_{n_2}$.}
    Each one of the $n_2$ tier-1 loops hosts $n_1$ data qubits (yellow circles) labelled as the elements $r^ks^l$ of $C_{n_1} s^l$ ($s^l$ is omitted on the label carried by each qubit). There is one ancilla site (squares) on the belt beside each data qubit. These tier-1 loops realise the $C_{n_1}$ factor, in the same fashion as \cref{fig:cyclic_group}. The loops are additionally coloured in a chequerboard pattern, differentiating a mode A (orange) and B (purple). Same-mode loops always have the same relative rotation direction (they can be controlled together) while different-mode loops alternatively co- or counter-rotate.
    The second factor of the group product is obtained by \textit{nesting}, or concatenating, these tier-1 loops into a higher-tier loop, via tier-2 shuttling tracks. By ordering the $C_{n_1} s^l$ groups in the clockwise direction as $l$ grows, one realises the $C_{n_2}$ factor. The faded thinner shuttling tracks are optional shortcuts that may shorten shuttling distances or be used to route around defects. Each tier-1 loop has an initialisation and measurement region by its side. In total the block holds $|G| = n_1 n_2$ data qubits.}
    \label{fig:group_embedding}
\end{figure}

The embedding on shuttling tracks is represented in \cref{fig:group_embedding}. Like for the cyclic group, we consider loops hosting static data qubits interfaced with shuttled ancillas. Qubits are now partitioned across $n_2$ loops of $n_1$ qubits, equidistantly placed on a $5\times (2n_2-3)$ square grid. Neighbouring hosting loops are separated by three empty loops that serve as buffer to avoid collisions during ancilla qubits transfer (see \cref{fig:motions}). A given hosting loop, say the $l$-th one, corresponds to the set of qubits $C_{n_1}s^l=\{r^ks^l\}_{0\leq k<n_1}$. Middle (corner) loops are indexed in the (anti)clockwise direction, with their origin $r^0s^l$ on distinct edges. This differentiation of middle and corner loops will be crucial for the synchronisation of the qubits motion. Loops are also divided into two groups that we denote mode A and B. Same-mode loops always have the same relative rotation direction (they can be controlled together) while different-mode loops alternatively co- or counter-rotate. Note that this concept is independent from the labelling of the static data qubits. Lastly, loops are connected by a network of shuttling tracks forming a square grid and permitting the transfer of qubits from loop to loop. Most edges may be removed from the grid to leave space for control electronics and measurement apparatuses. A minimal set of tracks forming a higher-tier loop connecting the hosting loops is represented. Additional tracks (narrower and faded in \cref{fig:group_embedding}) may be added to further reduce the shuttling distance or route around potential defects.

It already becomes apparent how taking the (semidirect) product of two cyclic groups naturally leads to the NeST architecture. The first factor $C_{n_1}$ of the product in \cref{eq:2_tier_group} requires qubits to move in a cyclic motion, which is simply implemented with a looped track as in \cref{fig:cyclic_group}. The second group $C_{n_2}$ adds another layer of nesting, or concatenation, where now $n_2$ groups of $n_1$ qubits are required to move in a cyclic motion. This leads to the embedding of $n_2$ looped tracks, arranged in a cyclic way, in other words forming a higher-tier loop. From now on and until the rest of the paper, we will describe the different loops according to their tier: the current scenario corresponds to a two-tier architecture, composed of $n_2$ tier-1 loops each hosting $n_1$ data qubits.

\subsubsection{Motions}

In order to understand the movement of the ancillas (remember these are the only ones that are being shuttled), we must compute the left and right group actions. Given an ancilla labelled $r^k s^l$, the result of the left or right multiplication of $r^k s^l$ with $r^i s^j$ represents the final destination of the ancilla.
\begin{align} 
    L(r^i s^j)r^k s^l &= r^{i+(-1)^jk} s^{j+l} \label{eq:left_mult} \\
    R(r^i s^j)r^k s^l &= r^{k+(-1)^li} s^{j+l} \label{eq:right_mult}
\end{align}
Let us first analyse the right multiplication case. The amount by which ancilla $r^k s^l$ must be displaced is $r^{(-1)^li} s^{j}$. Importantly, this quantity is independent of $k$ and the only $l$-dependence indicates the rotation direction of a given loop. This means that same-loop qubits are all shuttled by the same amount, either in the clockwise or anti-clockwise direction. Additionally, all qubits must be brought forward in the tier-2 structure by $j$ tier-1 loops. As for the left multiplication, the above equations tell us that for even $j$, the displacement takes the simple form $r^is^j$: all qubits must co-rotate by the same amount $i$, and hop over $j$ loops --- this is analogous to the previous case, except that even and odd loops were counter-rotating. For odd $j$, the destination now is $\propto r^{-k}$ rather than $r^k$. On top of the $r^is^j$ motion, all loops must therefore implement a reflection of their indices.

\begin{figure}
    \centering
    \includegraphics[width=\linewidth]{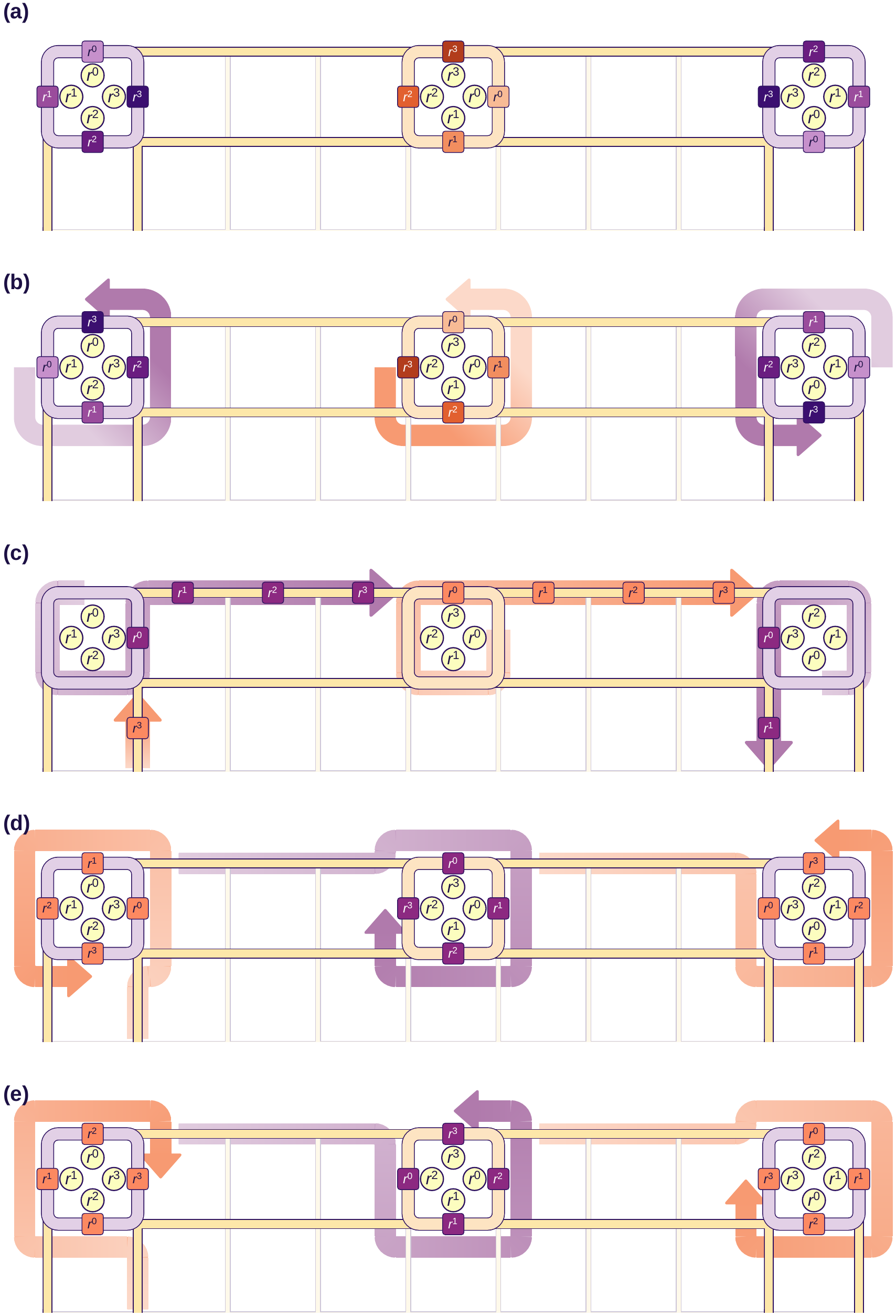}
    \caption{\textbf{Implementation of all motions required by $G = C_{n_1} \rtimes_{-1} C_{n_2}$.} Each panel shows the hosting loops $C_4$, $C_4 s$ and $C_4 s^2$ \textit{i.e.} the top row of \cref{fig:group_embedding}. \textbf{(a)} Initial configuration: ancilla and data labels are aligned. \textbf{(b)} Intra-loop rotation: each hosting loop rotates its $n_1$ ancillas collectively along its belt. Same-colour loops always follow the same relative rotation direction, but different-colour loops alternatively co- or counter-rotate (left or right multiplication by a rotation). Here a right multiplication by $r$ is shown: mode-A (-B) ancilla $r^i$ moves to the site of $r^{i+1}$ ($r^{i-1}$). \textbf{(c)} Inter-loop rotation, departure: the ancillas of each loop leave as a train for the next loop of the ring, riding the belt to an exit corner and then along the interconnecting tracks, from $C_4$ and $C_4 s$ along the outer track and from $C_4 s^2$ down the inner track. \textbf{(d) and (e)} Inter-loop rotation, arrival: $C_4 s$ now hosts the ancillas of $C_4$, and $C_4 s^2$ those of $C_4 s$. Trains can take either direction as they reach their destination loop (left or right multiplication by a reflection). If they enter the destination following the data qubits' alignment (panel (d)), a global rotation is implemented. If the other direction is taken (panel (e)) then a reflection is obtained.}
\label{fig:motions}
\end{figure}

The mapping of all the above motions onto NeST is represented in \cref{fig:motions}. Initially, ancilla and data qubits indices are aligned pairwise (panel (a)). Panel (b) illustrates the tier-1 rotations, which are partitioned into two modes: mode A (orange) and mode B (purple). As explained before, same-mode loops always have the same relative rotation direction, enabling their collective control, while different-mode loops alternatively co- or counter-rotate depending on the group action (left or right). This distinction is different from the data qubits ordering in each tier-1 loop, which was chosen to synchronise the inter-loop motions (see below).

The bottom three panels show the more global tier-2 movement: first qubits exit their hosting tier-1 loop like a train, travelling in lockstep (panel (c)); second they wrap around their destination loop, completing the global movement (panels (d) and (e)). Importantly, before initiating this last step, qubit trains can enter their final loop in either direction: one choice preserves their ordering with respect to the data qubits (up to a tier-1 loop rotation) while the other implements a reflection. For a loop hosting data qubits in, say, the clockwise direction, entering it in the same direction implements a global rotation (panel (d)), while following the anticlockwise direction implements a reflection (panel (e)). Note that the additional effect of this tier-2 motion is a tier-1 rotation by a quarter of a loop in the anticlockwise direction for global rotations (nothing for reflections). One can verify that the same applies to all loops, but that the loop-dependent indexation and labelling orientation of the data qubits is crucial to achieve this --- corner and middle loops do not follow the same rules. In other words, the labelling was chosen to abide by the guidelines we stated at the start of the section: global control (subgroups of $n_1$ qubits are shuttled collectively) and synchronisation. Using a different labelling could have introduced unwanted latencies.

The case $n_2=2$, which generates the dihedral group $D_{2n}$, helps gain intuition. $D_{2n}$ is the group of symmetries of an $n$-sided polygon --- $n$ rotations and $n$ reflections. What we have just shown is that NeST can implement those symmetries natively: rotations by going around tier-1 loops and reflections by entering the destination loop in the opposite direction. The restriction to these permutations only is intimately related to setting $srs^{-1}=r^{-1}$. In the outlook section, we will touch upon possible extensions to \textit{e.g.} $srs^{-1}=r^{a}$ where $a\neq-1$.  

A per-ancilla accounting of the four motions (\cref{app:2_tier_shuttling_upper}) gives a worst-case shuttling distance per stabiliser cycle of
\begin{equation} \label{eq:2_tier_shuttling_upper}
    L \leq \frac{\ell_{qq} n_1}{4} (2n_2+5),
\end{equation}
dominated by $n_1n_2\ell_{qq}/2$. This is about half the linear extent of a lattice holding $|G|$ qubits, hence comfortably inside the budget set by our first guideline.

Finally, note that there is a lot of flexibility regarding the arrangement of the tier-1 loops. For instance, the three-edge spacing between them is what keeps the uncurling trains of panel (c) from colliding. Other options exist: \cref{app:compact_embedding} shows a four-times more compact layout that recovers the same clearance by having the qubits zigzag, at the price of more intricate scheduling.

\subsection{Generalisation to $p$ cyclic groups} \label{sec:generalisation_p_groups}

Going further, one can envision a nested-loop hierarchy, embodying the (semidirect) product of $p$ cyclic groups (so far we have focused on $p=2$). Mathematically, the resulting group can be defined as:
\begin{equation} \label{eq:G_n_sigma}
    G(\vec{n},\sigma)=\big\langle r_1,\dots,r_p \;\big|\; r_i^{n_i}=1,\;\; r_i r_j r_i^{-1}=r_j^{\sigma_{ij}} \big\rangle
\end{equation}
where $\vec{n}=(n_1,\dots,n_p)$, the conjugation relations are imposed for $j<i$ only, and $\sigma=(\sigma_{ij})_{1\leq j<i\leq p}$ with
\begin{equation}
    \sigma_{ij}\in\{+1,-1\}\;\; \text{and} \;\; \sigma_{ij}=-1\Rightarrow n_i \text{ even}
\end{equation}
This simply describes the product of $p$ cyclic groups of generators $r_i$ and orders $n_i$. The commutation relation between generators $r_i$ and $r_j$ is specified by $\sigma_{ij}$ (commutes if $\sigma_{ij}=+1$, applies the inversion $r\mapsto r^{-1}$ otherwise).

Alternatively, we may refer to this group using the $\rtimes$ notation, in which case we will also specify the commutation relations. For example, a possible description of the group $G(\vec{n},\sigma)$ defined by $\vec{n}=(5,10,4)$, $\sigma_{21}=+1$, $\sigma_{31}=+1$ and $\sigma_{32}=-1$ may simply be:
\begin{equation*}
    G(\vec{n},\sigma) = C_{5} \times C_{10} \rtimes C_{4}
\end{equation*}
with the additional information: $s$ fixes $r$; $t$ inverts $s$ but fixes $r$ (where $(r,s,t)$ are the respective generators of $(C_{5}, C_{10}, C_{4})$). When a cyclic group fixes all previous letters, $\rtimes$ is replaced with $\times$. Every product symbol must be understood as acting on all previous groups (\textit{e.g.} we specify the action of $t$ on both $s$ and $r$) and the absence of brackets does not mean that the groups commute.

Note that a group is not generally described by a single pair $(\vec{n},\sigma)$. In the previous example, given that the first product is direct hence commutative, $\vec{n}=(10,5,4)$ and $(\sigma_{21},\sigma_{31},\sigma_{32})=(+,-,+)$ give another representation of the same group. Further, the number of factors can be varied, \textit{e.g.} with the identity $C_p \times C_q \cong C_{pq}$ for $\gcd(p,q)=1$: two commuting cyclic factors of coprime orders merge into one, generated by the product of their generators, provided every higher factor still acts on it by inversion or trivially (otherwise the group would escape our family). For instance, $(C_3 \rtimes C_2) \times C_5 \cong C_3 \rtimes C_{10}$, and $(C_7 \rtimes C_2) \times (C_5 \rtimes C_2) \cong (C_7 \rtimes C_{10}) \rtimes C_2$, where $C_{10}$ is generated by the product of the first reflection and the order-5 rotation.

\begin{figure*}
    \centering
    \includegraphics[width=\linewidth]{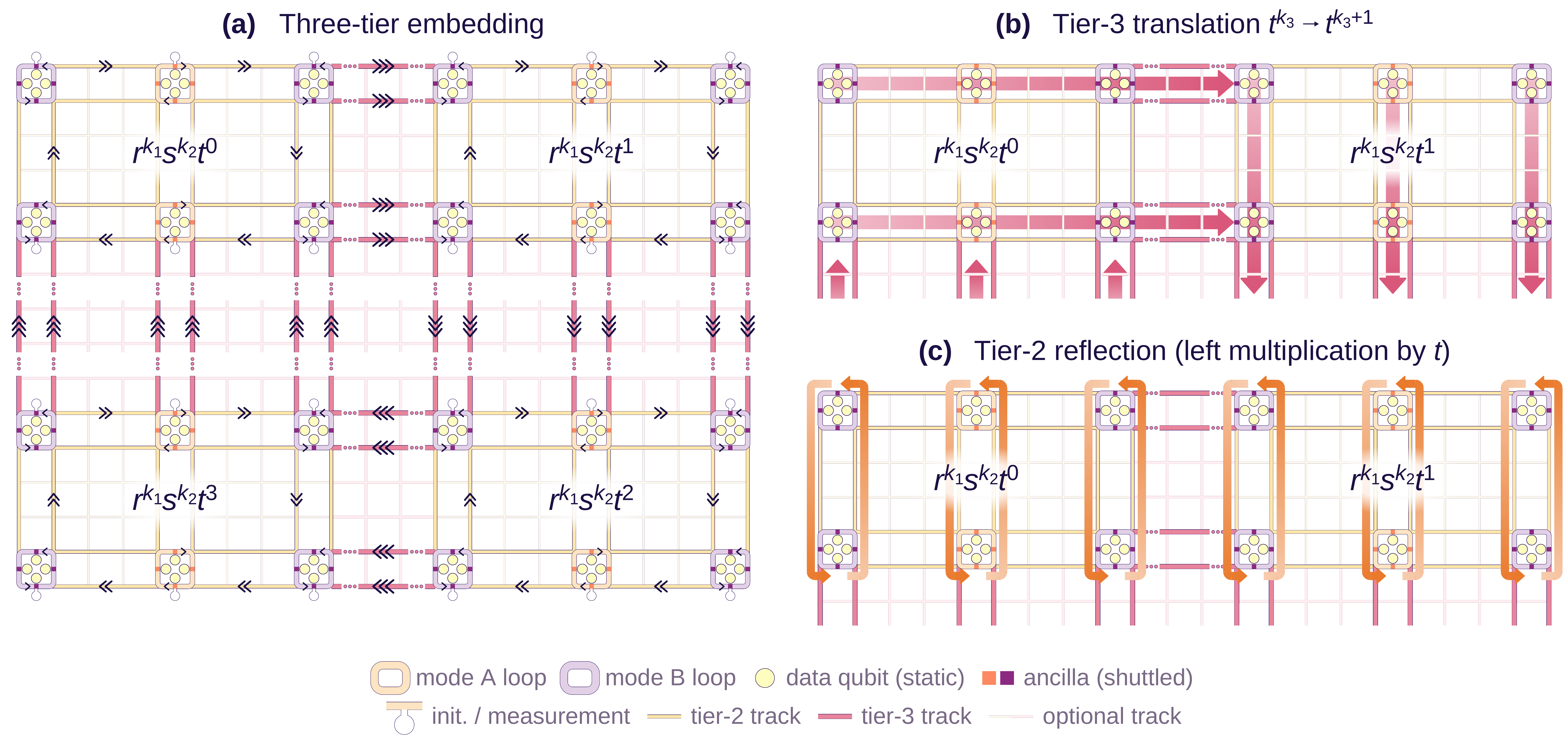}
    \caption{\textbf{Three-tier embedding and the motions it adds.} \textbf{(a)} Four tier-2 cells ($n_3=4$), each of them a copy of \cref{fig:group_embedding} (up to a few new non-optional tracks), are arranged as a $2\times(n_3/2)$ array and connected by tier-3 tracks (pink). Here $\vec{n}=(4,6,4)$. A cell is labelled by the tier-3 index it carries, $r^{k_1}s^{k_2}t^{k_3}$, and the cells are ordered clockwise in $k_3$ exactly as the tier-1 loops are ordered in $k_2$ one tier below. Left/right and top/bottom cells are spaced by $6$ and $2n_2-1$ edges respectively, chosen so that the trains of panel (b) never meet. Like for the lower tiers, the grid is strictly square and most tracks are optional (faded): they can be added to shorten routes or to steer around defects, or removed to free area for control electronics. \textbf{(b)} Tier-3 translation $t^{k_3}\to t^{k_3+1}$: each cell rides the pink tracks to the next position around the tier-3 ring, with every tier-1 loop inside it moving in lockstep, so the whole cell is displaced without any internal rearrangement. \textbf{(c)} Tier-2 reflection, required when $t$ inverts $s$ and obtained as a row swap inside each cell: top-row ancillas travel down and bottom-row ancillas travel up along the vertical tracks, which exchanges $C_{n_1}s^k$ with $C_{n_1}s^{n_2-1-k}$ and reverses the loop ordering. General tier-2 reflections follow by composing this move with tier-2 rotations. The same two primitives, translation around a tier-$k$ ring and reflection of the tier-$(k-1)$ indices, generate every motion at every higher tier.}
    \label{fig:3_tier_embedding}
\end{figure*}

In order to embed such a group with shuttling tracks, the idea is simply to extend the nesting philosophy we have been adopting to higher tiers. The two-tier architecture (\cref{fig:group_embedding}) was built by placing in a large loop $n_2$ of the cyclic group embeddings (\cref{fig:cyclic_group}); similarly a three-tier structure is built by interfacing $n_3$ two-tier embeddings. More specifically, these form a $2\times (n_3/2)$ array, plus interconnecting tracks. This is exactly shown in panel (a) of \cref{fig:3_tier_embedding}, where four tier-2 cells are connected. Pink shuttling tracks are the main tier-3 tracks, which allow elements from the tier-2 cells to travel in a cyclic motion. There is a spacing of 6 and $2n_2-1$ edges between the left/right and top/bottom tier-2 structures (chosen to avoid collisions in tier-3 movements). Like before, the shuttling grid is strictly square but not all tracks are required --- it may be made denser to accelerate the shuttling or route around potential defects, or sparser to leave space for control electronics. Note that the tracks between vertically-aligned tier-1 loops ($C_{n_1}s^k$ and $C_{n_1}s^{n_2-1-k}$) are now non-optional as this will help with the implementation of tier-2 reflections. The same principles extend to a four-tier layout, where tier-3 cells are now spaced by $2n_2+7$ and $n_2n_3-2$ edges in the left/right and top/bottom direction respectively. To leading order, the width of a $p$-tier architecture is $\sim 4\times n_2/2 \times\ldots\times n_p/2$ edges \textit{i.e.} a distance $\ell_{qq}|G|/2^{p-1}$. The vertical width is relatively smaller (no $n_p$ term).

The generalisation of the two-tier left and right multiplications (\cref{eq:left_mult} and \cref{eq:right_mult}) to a $p$-tier system is natural. Motions are decomposed into tiers, first implementing higher-tier rotations or reflections --- in the $p=2$ case we were also starting with the global movement before applying intra-loop rotations. Motions within tier-$k$ with $k\leq2$ are not modified and qubits belonging to distinct tier-2 structures (different values of $k_3$ in \cref{fig:3_tier_embedding}) can all move synchronously. Tier-3 motions correspond to cyclic permutations of the tier-2 groups, which can be achieved by translating them along the pink interconnecting shuttling tracks (\textit{e.g.} the group $r^{k_1}s^{k_2}t^{0}$ moves horizontally to the $r^{k_1}s^{k_2}t^{1}$ location and so forth). This is shown in panel (b) of \cref{fig:3_tier_embedding}. One subtlety arises when $t$ anticommutes with $s$, requiring tier-2 reflections --- the same way $(r,s)$ anticommutation required reflections of the tier-1 loops indices. Zooming on a tier-2 system (\textit{e.g.} that of \cref{fig:group_embedding}), the simplest tier-2 loops reflection is a row swap: top-row ancillas travel downwards while bottom-row ancillas travel upwards, using the vertical shuttling tracks like a racetrack (panel (c) of \cref{fig:3_tier_embedding}). This is why these tracks were here made non-optional. More general reflections are then decomposed as a product of this reflection and already-implementable tier-2 rotations. Higher-tier motions naturally follow from the above principles.

The architecture naturally abides by the last two guidelines we set at the start of the section. Subgroups of $n_1$ qubits initially in the same tier-1 loop are indeed always collectively shuttled, with larger subgroups of qubits also following preset patterns. The edge lengths are besides chosen to avoid clashes and synchronise all motions. As for the shuttling distance requirement, we can establish an upper bound on the per-move per-ancilla shuttling distance. To do so, we simply consider the longest possible move, which contains a multiplication by $r_p^{n_p/2}$ \textit{i.e.} shuttling between opposite corners of the tier-$p$ structure. To leading order, this move requires shuttling over the device width \textit{i.e.} $\ell_{qq}|G|/2^{p-1}$. Note how this term shrinks exponentially with the number of tiers. This proves the efficacy of our shuttling protocol, which finds shortest routes in an expanded, more complex lattice, and satisfies our shuttling distance criterion.

\subsection{Lifted product codes in spin qubit architectures} \label{sec:lifted_product_spin_arch}

So far, we demonstrated how the specific class of groups $G(\vec{n},\sigma)$ defined in \cref{eq:G_n_sigma} can efficiently be represented on a strictly square lattice of shuttling tracks. In other words, we laid out a protocol to implement one block of the total parity check matrix of \cref{eq:parity_check_mitten}. In reality, a code block is not embedded on a single structure like the one we studied in the previous section --- we will call it the unit cell --- but rather on multiple interconnected ones. This is represented in \cref{fig:main_figure}, in the specific case of five two-tier unit cells defined by $\vec{n}=(8,4)$. Like before, ancilla qubits are free to roam each cell, but can additionally travel between cells through the dark purple shuttling highways, enabling any long-range connectivity required by the base codes $A$ and $B$. The embedding still respects the square lattice geometry and the 2D arrangement of the cells directly reproduces the protograph of the lifted-product code. If a schedule is found so as to minimise the shuttling distance and avoid all ancilla qubits collisions, it becomes theoretically possible to implement any lifted-product code based on the group $G(\vec{n},\sigma)$. This includes every quasi-cyclic code~\cite{panteleev_kalachev_lifted_2021, bravyi_bb_2024} --- indeed, setting $\sigma\equiv+1$ makes $G(\vec{n},\sigma)$ the direct product $C_{n_1}\times\cdots\times C_{n_p}$, and every finite abelian group is of that form, so the whole abelian lifted-product family is implementable (with notable $30\%$-rate codes from \cite{cain2026shorsalgorithmpossible10000} with distance up to 24). On the non-abelian side, certain 2BGA codes \cite{2BGA_codes} \textit{e.g.} $\llbracket 96,10,12 \rrbracket$ and some very promising mitten codes \cite{bhardwaj2026highrateqldpcprocessors} are implementable as well (see \cref{sec:perf_eval}). \cref{fig:main_figure} represents the latter codes.

An inter-cell transfer may be implemented by simply translating an entire cell through the shuttling highways onto the destination cell, which to leading order, is about twice the cell width \textit{i.e.} $\ell_{qq}|G|/2^{p-2}$. We do not delve into a detailed scheduling as it is rather code-specific, even gate-ordering specific. We may simply estimate the total shuttling distance for a mitten code. For this family, a given ancilla must visit two corners plus the central cell, and
perform three interactions in each, for a total stabiliser weight $w=9$. As we will see in \cref{sec:circuit_implementation}, it is preferable to split in two one block's interactions to avoid circuit distance reductions, meaning three rather than two inter-cell switches. Adding up all contributions, one may find a leading-order estimate of the per-ancilla shuttling distance over an entire stabiliser cycle:
\begin{equation*}
    L \approx 9 \times \ell_{qq}|G|/2^{p-1} + 3 \times \ell_{qq}|G|/2^{p-2}
\end{equation*}
Using $n=5|G|$ for mitten codes, one obtains:
\begin{equation} \label{eq:max_shuttling}
    L \approx \frac{1}{3\times2^{p-1}}\,n\,w\,\ell_{qq}
\end{equation}
As expected, this distance scales with the code size $n$ and the stabiliser weight $w$, but satisfies our shuttling distance requirement since the prefactor is strictly lower than 1. Note that this estimate overlooks any term that is not a product of all $n_i$'s, but also ignores further improvements a dedicated route planner can achieve.

\section{Performance evaluation} \label{sec:perf_eval}

So far, we have demonstrated how a simple square lattice can enable the efficient embedding of certain families of lifted-product codes, including all quasi-cyclic codes and some 2BGA and mitten code constructions \cite{panteleev_kalachev_lifted_2021, bravyi_bb_2024, 2BGA_codes, bhardwaj2026highrateqldpcprocessors, cain2026shorsalgorithmpossible10000}. In this section, we focus on the mitten family and estimate its performance on our proposed embedding via logical error rate simulations. On top of their very attractive $20\%$ rate and fully worked-out logic, we choose these codes as they are based on a very simple protograph, thus do not require complicated inter-cell scheduling. Indeed, most of the past section was spent building the architecture and optimising the intra-cell motions, leaving inter-cell movements on large protographs as an open problem. One particular feature we wish to observe is the tolerance of these codes to shuttling errors, which will represent a growing proportion of the total injected noise as the code size increases.

\begin{figure}
    \centering
    \includegraphics[width=\linewidth]{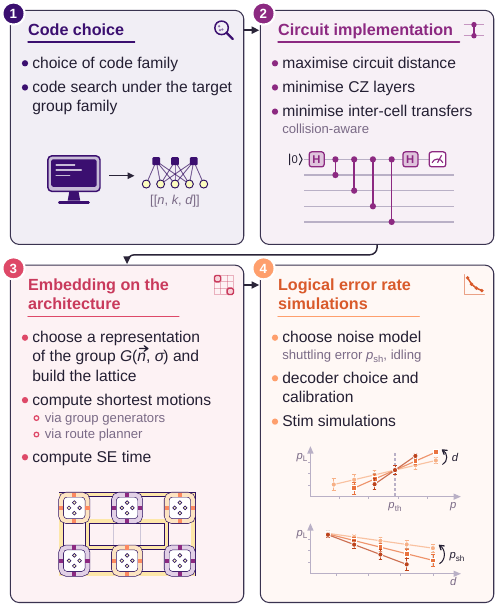}
\caption{\textbf{The four stages of the simulation pipeline}, each corresponding to one subsection of \cref{sec:perf_eval}. \textbf{(1)} Code choice (\cref{sec:codes}): a code family is fixed --- here the mitten codes --- and a deep search returns $\llbracket n,k,d\rrbracket$ codes whose base group lies in the implementable family $G(\vec{n},\sigma)$. \textbf{(2)} Circuit implementation (\cref{sec:circuit_implementation}): the order in which each ancilla block performs its nine CZ gates is optimised for circuit distance, subject to minimising the number of CZ layers and of inter-cell transfers, and to never placing two ancilla blocks in the same cell at once. \textbf{(3)} Embedding on the architecture: a group representation $(\vec{n},\sigma)$ is chosen, the lattice built, and every ancilla motion of the schedule routed --- either by decomposing the transition into tier generators or with a group-agnostic route planner --- which fixes the shuttling distances and hence the syndrome-extraction time. \textbf{(4)} Logical error rate simulations: the noise model is applied to the resulting Stim circuit, which is decoded and sampled to give the logical error rate as a function of physical error rate $p$, shuttling dephasing $p_\mathrm{sh}$ and code distance $d$.}
    \label{fig:simulation_pipeline}
\end{figure}

We here focus on memory, while logic will be addressed in \cref{sec:logic}. The simulation pipeline is summarised in \cref{fig:simulation_pipeline} and detailed in the following subsections.

\subsection{Choosing the codes} \label{sec:codes}

In the previous section, we demonstrated that not all, but specific classes of mitten codes could be implemented efficiently with our architecture: those based on the group $G(\vec{n},\sigma)$ defined in \cref{eq:G_n_sigma}. Of all flagship codes presented in \cite{bhardwaj2026highrateqldpcprocessors} (Table I), only the $\llbracket200,40,12\rrbracket$ built upon $C_4 \times D_{10}=C_4 \times (C_5\rtimes C_2)$ directly reads as implementable. One can however show that the base groups of the $\llbracket150,30,10\rrbracket$ and $\llbracket300,60,14\rrbracket$ codes may also be rewritten in the $G(\vec{n},\sigma)$ form:
\begin{align*}
    C_5 \times S_3 &= C_3 \rtimes C_{10} \\
    C_{10} \times S_3 &= C_6 \rtimes C_{10}
\end{align*}
On the contrary, no higher-distance code from that paper satisfies this desirable property. In order to bridge this gap, we therefore initiated a deep code search aiming at finding good mitten codes built upon our target group class. Theorem~2 of Ref.~\cite{bhardwaj2026highrateqldpcprocessors} upper-bounds the mitten code distance $d$ by the distances of the two base codes. Therefore the workflow consisted in finding good classical base codes first and before evaluating the parameters of the quantum code they would generate. Technical details can be found in \cref{app:code_search}. The results of the code search led to the four bottom codes shown in \cref{tab:target_codes} (the top three ones are among flagship codes of \cite{bhardwaj2026highrateqldpcprocessors}). A fuller description of the codes is given in \cref{tab:target_codes_full} of \cref{app:codes}. Crucially, this search found no evidence that restricting to $G(\vec{n},\sigma)$ caps the achievable distance, and returned candidates large enough to run deep quantum algorithms (see \cref{sec:ler_sims}). Our codes also offer similar parameters as same-distance ones given in Ref.~\cite{bhardwaj2026highrateqldpcprocessors} (slightly smaller block size for distance 16, and slightly larger for distance 18 and 20).

\begin{table}[htbp]
  \centering
  \begin{tabular}{llcc}
    \toprule
    $\llbracket n,k,d \rrbracket$ & Group & $d_{\mathrm{circ}}$ & SE time \\
    \midrule
    $\llbracket 150,30,10 \rrbracket$ & $C_{3} \rtimes C_{10}$ & $\leq 8$ & $16.4~\mu$s \\
    $\llbracket 200,40,12 \rrbracket$ & $C_{5} \times C_{4} \rtimes C_{2}$ & $\leq 10$ & $17.5~\mu$s \\
    $\llbracket 300,60,14 \rrbracket$ & $C_{6} \rtimes C_{10}$ & $\leq 13$ & $20.7~\mu$s \\
    $\llbracket 480,96,\leq 16 \rrbracket$ & $C_{3} \times C_{4} \rtimes C_{4} \rtimes C_{2}$ & $\leq 15$ & $26.1~\mu$s \\
    $\llbracket 600,120,\leq 18 \rrbracket$ & $C_{5} \rtimes C_{6} \rtimes C_{4}$ & $\leq 17$ & $29.3~\mu$s \\
    $\llbracket 700,140,\leq 20 \rrbracket$ & $C_{7} \rtimes C_{10} \rtimes C_{2}$ & $\leq 20$ & $30.1~\mu$s \\
    $\llbracket 1400,280,\leq 28 \rrbracket$ & $C_{7} \rtimes C_{10} \rtimes C_{4}$ & --- & $\approx 42~\mu$s$^{\dagger}$ \\
    \bottomrule
  \end{tabular}
  \caption{
  Target mitten codes built upon the implementable group $G(\vec n,\sigma)$. The first three codes are directly taken from Ref.~\cite{bhardwaj2026highrateqldpcprocessors} (with the base groups rewritten to match the form of $G(\vec n,\sigma)$). They have exact distances. The last four codes are obtained with a deep code search, with distances upper-bounded by $10^{8}$ sQetch trials. The full description of the group (commutation relations between cyclic group generators) and of the codes (entries of the base codes) can be found in \cref{tab:target_codes_full}. $d_{\mathrm{circ}}$ is the circuit-level distance of the syndrome-extraction schedule, evaluated by an information-set search over circuit faults, including the hook errors of the chosen CZ ordering; each value is an upper bound, certified by a logical fault pattern of that weight injected into the Stim circuit. The last column is the full syndrome extraction time ($X$ then $Z$ stabiliser measurement), using $200$~ns CZ layers, $0.5~\mu$s single-qubit gates, $1~\mu$s readout and shuttling at $v=50$~m/s. $^{\dagger}$Worst-case estimate without a gate ordering, as no circuit-level distance was evaluated for this code.
  }
  \label{tab:target_codes}
\end{table}

\subsection{Circuit implementation} \label{sec:circuit_implementation}

Once the codes are fixed, the next step consists in devising their stabiliser circuit implementation. We choose to focus on a simpler circuit where $X$ and $Z$ stabilisers are not interleaved. Ancilla initialisation and measurement happen in the $Z$ basis. Entangling gates are chosen to be CZ gates, which are native to silicon spin qubits \cite{meunier_cz, Burkard_cz, Xue_2022}. Hadamard gates (also native \cite{Xue_2022, Noiri_2022, vahapoglu_single-electron_2021}) are inserted after ancilla initialisations, before ancilla measurements, and on all data qubits at the start and the end of the $X$ stabiliser layer.  While direct CNOTs could also be used in place of the H-CZ-H sequence \cite{Noiri_2022, Mills_2022}, the combined use of CZs and the `$X$ then $Z$' stabiliser circuits enables an additional trick that helps reduce idling errors, as we will explain in \cref{sec:noise_models}. Besides, since shuttling errors are mostly $Z$-like, using CZs prevents error propagations to the data qubits, meaning that those errors only ever become syndrome measurement errors, which are the less damaging kind.

Regarding the gate ordering, we keep a few criteria in mind: $(i)$ achieving a high circuit distance; $(ii)$ minimising the number of CZ layers (at least 9 per stabiliser type) and $(iii)$ minimising the overall shuttling. This latter requirement aims at reducing shuttling errors and accelerating the stabiliser cycle. In this step, we do not specify the structure of the unit cells just yet, but rather consider a lattice of five cells which can only host one ancilla block at a time. In particular, we wish to minimise the number of inter-cell transfers, \textit{i.e.} implement interactions from the same matrix block in \cref{eq:parity_check_mitten} at once if possible. This avoids unnecessary round trips for the ancilla qubits. For a mitten code, each ancilla block must visit three unit cells (two corners plus the centre), which amounts to at least two inter-cell shuttles. We additionally guarantee no collision at this stage, meaning that no two ancilla blocks can occupy the same unit cell at the same time. Details on the gate ordering optimiser can be found in \cref{app:gate_ordering}.

This study showed that for most codes, restricting ourselves to two inter-cell switches (the minimum) had detrimental effects on the circuit distance, while accepting three switches enabled circuit distances $d_\text{circ}$ upper bounds close to the code distance $d$ upper bounds ($d-2$ for our first two codes, $d-1$ for the next three and exactly $d$ for the $\llbracket 700,140,\leq 20\rrbracket$ code, see \cref{tab:target_codes}). No circuit distance was computed for the $\llbracket 1400,280,\leq 28\rrbracket$ code as this proved computationally out of reach. Notably, the $\llbracket 480,96,\leq 16 \rrbracket$ code is the only code reaching $d_\text{circ}=d-1$ (or rather, the upper bounds reach this equality) with two inter-cell switches only.

\subsection{Embedding on the architecture} \label{sec:embedding}

Until now, the details of the unit cells have not been specified. The next step therefore consists in precisely mapping the code onto the NeST architecture and computing all the shuttling moves.

First, one must choose the group embedding $G(\vec{n},\sigma)$. Indeed, while the base group is fixed from the code itself, there may be multiple choices of the pair $(\vec{n},\sigma)$ representing it. As we alluded to in \cref{sec:generalisation_p_groups}, commuting cyclic groups may be swapped, and co-prime ones may be factorised together. We exploit this to find, for each code, a group description satisfying:
\begin{equation} \label{eq:ni_even}
    i\geq2 ~\Rightarrow~ n_i~ \text{even}
\end{equation}
This is simply because for $i>1$, we always designed the tier-$i$ layout as a $2\times (n_i/2)$ array of tier-$(i-1)$ structures, assuming even $n_i$. There is no objection however in having $n_1$ odd, \textit{i.e.} an odd number of qubits per hosting loops. The base group of each target code is given in \cref{tab:target_codes} in the form $C_{n_1} \rtimes \ldots \rtimes C_{n_p}$ \textit{i.e.} in increasing tier order, such that only the first factor may have an odd order. For instance, the $\llbracket 600,120,\leq 18 \rrbracket$ code is based on a three-tier unit cell defined by $\vec{n}=(5,6,4)$ (it is the unit cell of \cref{fig:3_tier_embedding} except that the small loops host five rather than four qubits).

Note that \cref{eq:ni_even} is only enforced for simplicity --- one may extend the layouts of \cref{sec:architecture} to odd $n_i$'s without much trouble. This would offer a richer space for optimisation, where each group representation would be attached to a given lattice structure with its own shuttling distances.

Once the layout is set, the second step consists in computing the ancillas motions, determined transition by transition for the chosen gate ordering. Between two consecutive CZ layers performed in the same unit cell, an ancilla labelled $x$ must be brought from $x$ to $h_L\, x\, h_R$, where $h_L$ and $h_R$ are fixed by the two interactions. The same permutation of the ancillas can be realised in several ways, and we compare a few of them. The default route decomposes $h_L$ and then $h_R$ into tier generators, and applies the collective motion of each generator one after the other as described in \cref{sec:architecture}. This is often longer than necessary, as the left and right parts of a transition can partially cancel. For instance, in $C_3 \rtimes C_{10}$, $s^2$ commutes with every element, so that $s^4 x s^4 = x s^8 = x s^{-2}$: instead of shifting all ancillas by four loops forward twice, a single shift by two loops backward yields the same result. To find such shortcuts, we search over all products of up to three collective motions for the shortest one with the same net effect as the transition. This approach is compared to a group-agnostic general-purpose route planner that looks for a single move bringing every group directly from its initial to its final loop, avoiding collisions between qubits. Among the valid candidates, we keep the one with the shortest mean distance per ancilla. When an ancilla block changes unit cell, all its loops are translated together to the same loops of the destination cell along the inter-cell highways (see \cref{fig:main_figure}(a)). Initialisation and measurement stations are placed on every unit cell, such that negligible additional shuttling is required before measurement. Note that the motions are optimised for a fixed gate ordering for simplicity, but optimising both jointly could further reduce the shuttling distance.

Once the full schedule has been computed, one can eventually deduce the stabiliser cycle times of each code, which are given in \cref{tab:target_codes} and all under the promised $50\mu$s mark (the last one estimated rather than routed, see below). Larger codes require longer stabiliser cycles due to increasing ancilla shuttling. As we did not search for a gate ordering for the $\llbracket 1400,280,\leq 28\rrbracket$ code, the syndrome-extraction time is only an estimate, obtained by applying the worst-case shuttling cost of a large subset of schedules on its $\vec{n}=(7,10,4)$ layout, rather than by routing a predetermined ordering.

These values are based on the timing of each operation in the stabiliser cycle: 200-ns CZ gates~\cite{Xue_2022, Noiri_2022}, 500-ns Hadamards~\cite{vahapoglu_single-electron_2021}, 1 $\mu$s for measurement and initialisation altogether~\cite{nurizzo_complete_2023} and a shuttling speed $v=50$m/s~\cite{desmet2024highfidelitysinglespinshuttlingsilicon}. All these values have already been demonstrated experimentally thus are not particularly conservative. They are summarised in \cref{tab:ops_params}.

\subsection{Logical error rate simulations} \label{sec:ler_sims}

We now have all the necessary elements to estimate the performance of the NeST architecture. To do so, we use Stim \cite{gidney2021stim} and run Monte-Carlo simulations for each code in both the $X$ and $Z$ bases, for $d$ rounds of syndrome extraction.

\subsubsection{Noise model} \label{sec:noise_models}

We adopt a circuit-level depolarising noise model as our main noise model. This choice is deliberate: different silicon spin qubit platforms implement their gates in appreciably different ways --- exchange-based versus resonantly driven, electron versus hole spins, varying magnetic field gradients --- so a model built around one particular gate implementation would tie our conclusions to it. A uniform depolarising model instead gives a hardware-agnostic baseline. It is also the conservative choice: in \cref{app:spin_qubit_noise_model} we construct a noise model specific to single-electron silicon spin qubits, in which gate infidelities and shuttling errors both descend from uncalibrated $g$-factor fluctuations. We find that all codes perform \emph{better} under it than under depolarising noise at the same average infidelity. This is a relatively general result: depolarising channels tend to act like a worst-case assumption since their noise has no structure the decoder can exploit.

Initialisation, measurement and single- and two-qubit gates errors are all set by a common error rate $p$. Shuttling errors are modelled with a dephasing channel of rate $p_\text{sh}\times L$ where $L$ is the shuttling distance. This describes uncalibrated variations of the $g$-factor landscape an electron travels through, which is the main source of shuttling noise \cite{losert_shuttling_2024, volmer_valley_mapping_2024, Langrock_2023}. Idling is dominated by $1/f$ noise, such that its mostly-quasi-static noise can be modelled by
\begin{equation} \label{eq:idle}
    p_\text{idle} = \frac{1}{2}\left(1-\mathrm{e}^{-(t/T_2^*)^2}\right)
\end{equation}
where $t$ is the idling time and $T_2^*$ the qubit's dephasing time. We fix it at $T_2^*=20\mu$s, which has largely been achieved today \cite{Struck_2020, yonedaQuantumdotSpinQubit2018}. Ancilla qubits are rarely idle during the course of the stabiliser cycle --- they are mostly shuttled --- but data qubits are. In fact, they spend at least $80\%$ of the stabiliser cycle time idling, as much as $28\mu$s for the largest code with a fully routed schedule, the $\llbracket 700,140,\leq 20 \rrbracket$. Adding up the contributions of all idling windows, the total idling error would be $p_\text{idle}=31\%$, which is far above threshold. Rather, we decide to Hahn-echo the data qubits \cite{hahn_spin_1950}: since idling noise is dominated by quasi-static detunings $\delta_q$, a wait of duration $t$ applies the coherent rotation $R_z(\delta_q t)=\mathrm{e}^{-\mathrm{i}\delta_q t Z/2}$. This can be refocused with an $X$ pulse placed in the middle of the idling window, using the identity $R_z(\theta)XR_z(\theta)=X$ for any $\theta$. The leftover $X$ is absorbed into the Pauli frame. Rather than applying this refocusing method between every CZ layer, we note that our choice of `$X$ then $Z$' stabiliser circuits combined with the use of CZ gates enables an additional trick. Namely, since $R_z$ commutes with CZ, all the waits of a qubit between two of its Hadamards add up to a single rotation $R_z(\delta_q W_q)$, with $W_q$ its total idle time in that window. Measuring $X$ and $Z$ stabilisers in separate rounds leaves each data qubit with only two windows per cycle (between Hadamard layers at the start and end of the $X$ stabiliser round). When the $X$ pulse cannot  be placed exactly at the idle time midpoint (because it would overlap with a CZ gate), we verify that the residual error is small by quantifying the non-refocused time. One may think that this scheme would break down in the presence of noisy non-diagonal CZs. However we show in \cref{app:hahn_echo} that the echo still cancels the full idle phase, and a small fraction of each CZ's error merely changes Pauli type at unchanged probability. We also neglect additional idling errors beyond quasi-static noise, set by $T_{2,\text{Hahn}}$ --- this parameter can be in the milliseconds timescale~\cite{burkard_2023}, much larger the microseconds-long idling windows. Ultimately, data qubits idling errors thus reduce to $p_\text{Hahn}=p$ (error due to the refocusing $X$ pulse) applied twice per stabiliser cycle.

For comparison purposes, we will also evaluate surface-code logical error rates. They will be implemented with CZ-based interleaved-$X$-then-$Z$ stabiliser circuits. Their noise model is the same as for the mitten codes, except that no shuttling noise is included. This assumes a dense grid of qubits with nearest-neighbour connectivity, which is somewhat optimistic (in favour of the surface code): given the small size of a quantum dot (100nm), spin-qubit-based architectures \cite{li_crossbar_2018, boterSpiderwebArraySparse2022, pataki_surface_code_crossbar_2024, kunne_spinbus_2024, escofet2025quantumreversemappingsynthesizing, cai_looped_pipelines_2023, siegel_two_by_n_2024, micciche2025optimizingcompilationerrorcorrection, siegel2025snakesplanemobilelow} always space qubits out to leave room for the control electronics, which then requires some amount of shuttling.

\begin{table}[]
    \centering
    \begin{tabular}{cccc}
        \toprule
        Operation & Duration & Error rate & Noise structure \\
         \hline
        CZ & 200ns & $p$ & Depolarising \\
        H & 500ns & $p$ & Depolarising \\
        Measurement & 1$\mu$s & $p$ & Bit-flip \\
        Initialisation & N/A & $p$ & Bit-flip \\
        Shuttling (per $\mu$m) & 20ns & $p_\text{sh}\times L$ & Dephasing \\
        Idling (ancilla) & N/A & $p_\text{idle}$ & Dephasing \\
        Idling (data) & N/A & $p_\text{Hahn}$ & Depolarising \\
        \bottomrule
    \end{tabular}
    \caption{\textbf{Operation durations and noise channels} used throughout the simulations. A single parameter $p$ sets the error rate of every gate, initialisation and measurement; $p_\mathrm{sh}$ is the dephasing incurred per micrometre of shuttling, so that a move of length $L$ (carried out at a speed $v=50$m/s) injects an error $p_\mathrm{sh}\times L$. Initialisation time is included in the measurement, since a qubit is reset by the measurement that precedes it. The idling error $p_\mathrm{idle}$ of ancilla qubits is set by $T_2^*=20\mu$s. Data qubits do not accumulate this quasi-static noise as they are rather Hahn-echoed: their idling cost reduces to the error $p_\mathrm{Hahn}=p$ of the refocusing pulse, applied twice per stabiliser cycle. All durations are at or within reach of experimentally-demonstrated performance.}
    \label{tab:ops_params}
\end{table}

\subsubsection{Decoder}

Decoding is performed with the telescoping decoder of Ref.~\cite{bhardwaj2026highrateqldpcprocessors}, which runs a sequence of increasingly expensive stages and passes a shot on to the next stage only when the current one fails to return a trusted correction. Its first stages are ensembles of belief-propagation decoders, which handle the vast majority of shots, while the last stage looks for the most likely error consistent with the syndrome. In Ref.~\cite{bhardwaj2026highrateqldpcprocessors}, this last stage is an integer program solved with the commercial solver Gurobi. We replace it with the open-source search decoder Tesseract~\cite{beni2025tesseract}, keeping a BP+OSD correction as a fallback whenever it is lighter. Below threshold, this stage barely affects the runtime: at $p=0.1\%$, at most $0.07\%$ of shots reach it and it takes under $5\%$ of the decoding time for every simulated code, so the runtime is set by the belief-propagation stages. We additionally checked that Tesseract successfully decodes the shots the belief-propagation stages could not handle. As a check, decoding the $\llbracket 150,30,10 \rrbracket$ code under the noise model and gate ordering of Ref.~\cite{bhardwaj2026highrateqldpcprocessors} closely matches their memory logical error rates. \red{}

\subsubsection{Monte-Carlo simulations}

\begin{figure}
    \centering
    \includegraphics[width=\linewidth]{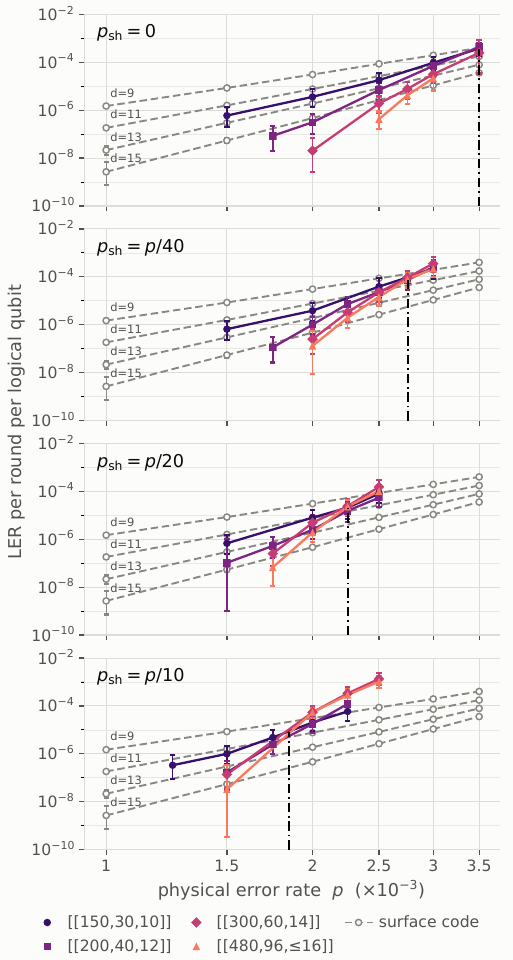}
    \caption{\textbf{Logical error rate per round per logical qubit against the physical error rate $p$}, for the first four codes of \cref{tab:target_codes} compiled onto the NeST architecture and simulated for $d$ rounds of syndrome extraction under depolarising noise. The four panels differ only in the shuttling dephasing rate, from $p_\mathrm{sh}=0$ (top) to $p_\mathrm{sh}=p/10$ (bottom). The third panel, $p_\mathrm{sh}=p/20$, is the operating point targeted in the text, a factor ten below the fidelity demonstrated in Ref.~\cite{desmet2024highfidelitysinglespinshuttlingsilicon}. Dashed grey lines are surface codes of distance $9$ to $15$ under the same noise model but with no shuttling error, which is conservative in their favour. The vertical dash-dotted line in each panel marks the threshold; it stays above $p=0.1\%$ throughout.}
    \label{fig:ler_vs_p_depol}
\end{figure}

Results from the simulations are presented in \cref{fig:ler_vs_p_depol}, where the logical error rate per round per logical qubit is plotted against the physical error rate $p$. Panels from top to bottom correspond to increasing levels of shuttling dephasing, the top one being dephasing-free. For reference, state-of-the-art shuttling achieved a fidelity of $99.95\%$ per micrometer \cite{desmet2024highfidelitysinglespinshuttlingsilicon}. For a $0.1\%$ depolarising machine this corresponds to $p_\text{sh}=p/2$, five times above the worst-case we plotted. The coloured lines correspond to the first four target codes of \cref{tab:target_codes}, evaluated under the constraints of the NeST architecture (the last three were not simulated due to numerical limitations). The vertical dash-dotted black line in each panel marks the threshold. We compare them to similar-distance surface codes under the same noise model but with no shuttling error (see the last paragraph of \cref{sec:noise_models}). The same simulations run under the silicon-spin-qubit-specific noise model of \cref{app:spin_qubit_noise_model} are given in \cref{fig:ler_vs_p_tailored} --- they sit below the curves shown here, confirming that the depolarising model is the more pessimistic of the two.

The first observation from the figure is that for any plotted level of shuttling dephasing, the threshold of the mitten codes is above $p=0.1\%$, which is a common target error rate for a utility-scale machine: at this working point increasing the code size reduces the logical error rate (slightly sub-exponentially, as we will note at the end of this section). As expected, the threshold is lowered by higher values of $p_\text{sh}$.

One can also note how the mitten codes performance is on par with or better than the surface codes for every shuttling assumption up to $p_\mathrm{sh}=p/20$, despite receiving much more physical noise (nine CZ gates per qubit, shuttling noise and more idling errors). They however achieve this at a much higher code rate: $20\%$ instead of $1/d^2$. The gradients of the mitten codes are also much steeper than same-distance surface codes: this manifests the fact that the mitten codes only have few distance-$d_\text{circ}$ undetectable error patterns. Therefore a scaling with $p^{d_\text{circ}/2}$ would only be observed at much lower physical error rates than plotted. This was already observed in Ref. \cite{bhardwaj2026highrateqldpcprocessors}.

\begin{figure}
    \centering
    \includegraphics[width=\linewidth]{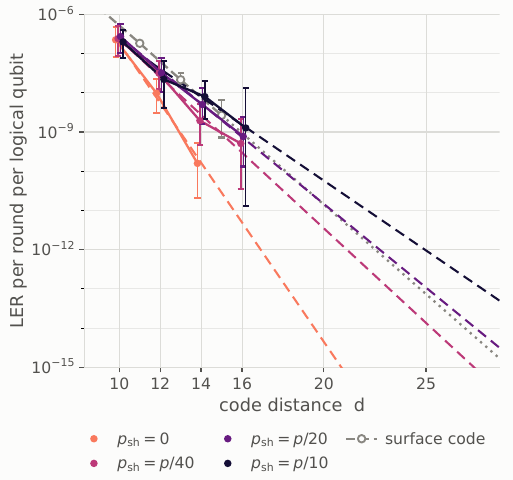}
    \caption{\textbf{Extrapolation to a utility-scale machine at $p=0.1\%$.} Points are the logical error rate per round per logical qubit of the four simulated codes, obtained by fitting $y=ap^{d_\mathrm{circ}/2}$ from the lowest physical error rate reachable for all codes by Monte-Carlo sampling --- deliberately more conservative than extrapolating the measured slope, which is steeper. Colours distinguish the four shuttling assumptions; the grey dotted line is the surface code with no shuttling noise. Dashed lines are linear regressions in log scale, extended to distance $27$. At the targeted $p_\mathrm{sh}=p/20$, a distance-27 mitten code would reach a per-logical per-round error of $10^{-14}$ for $9$ physical qubits per logical qubit, which the $\llbracket 1400,280,\leq 28\rrbracket$ code of \cref{tab:target_codes} may attain. The extrapolation holds the shuttling noise fixed, whereas it in fact grows with the block size; see the caveats in the text.}
    \label{fig:ler_vs_d_depol}
\end{figure}

Further, we wish to estimate the performance of the codes in a utility-scale $p=0.1\%$ machine, which is represented in \cref{fig:ler_vs_d_depol}. For each shuttling dephasing noise assumption, data points are obtained from fitting the form $y= ap^{d_\text{circ}/2}$ from the lowest physical error rate that could be accessed via the Monte-Carlo simulations to all codes. This is more conservative than directly extrapolating the data --- which has a steeper slope than $d_\text{circ}/2$ as explained above. Data points are then fitted with a linear regression in log scale. Targeting a per-logical per-round error of $10^{-14}$ \cite{gidney_ekera_2021}, our $\llbracket 1400,280,\leq 28\rrbracket$ code may suffice under $p_\text{sh}=p/20=0.005\%$ per micrometre, one order of magnitude better than demonstrated shuttling fidelities. This is $(2n-k)/k=2\times5-1=9$ physical qubits (data and ancilla) per logical, while a surface-code implementation would also require a distance $d_\text{SC}=27$ but $2d_\text{SC}^2-1=1457$ physical qubits per logical. This is more than 100x compression.

A few caveats regarding this analysis: first this is only a very rough estimate given the relatively large error bars on the data; second this assumes that the code distance of $\llbracket 1400,280,\leq 28\rrbracket$ does reach the established upper bound, and that its circuit distance can closely match it as well; third and most importantly, this ignores the fact that the noise here scales with the code size. Indeed, the shuttling distance, thus its associated noise, grows as the codes do. This means that the logical-error-rate-versus-$d$ predictions should actually curve up as the distance increases, rather being a strict straight line. This effect is however mitigated by our design choices: since we only shuttle the ancillas and run CZ-based stabiliser circuits, shuttling dephasing errors never spread to the data qubits and only ever become measurement errors, which are the least damaging kind.

\section{Towards logic} \label{sec:logic}

So far, we have only considered quantum memory. A fault-tolerant processor must also compute, and on qLDPC codes this is usually achieved by measuring logical Pauli operators through code surgery. During this process, a graph surgery gadget is attached to the code, and the question for NeST is whether the new deformed code can also be operated with rigid, globally controlled motions. After recalling the relevant notions, we show that such constructions exist for mitten codes, at a modest cost in gate layers and none in qubits.

\subsection{Graph surgery with mitten codes}

Universal quantum computation can be carried out with logical Pauli-product measurements supplemented by magic states. On qLDPC codes, these measurements are typically realised by code surgery~\cite{horsman2012latticesurgery, cohen2022lowoverhead, cross2024improvedqldpcsurgery, williamson2026gauging, ide2025homological, swaroop2024adapters, he2025extractors}: an ancillary system, the \textit{gadget}, is temporarily attached to the code block, forming a larger \textit{merged code} in which the target logical operator becomes a product of stabilisers. Measuring the merged code for $O(d)$ rounds reads out the logical value reliably, after which the gadget is detached and the original code restored.

Let us consider the measurement of an $X$-type logical operator $\bar X$ of support $S$, following the graph-surgery construction of Ref.~\cite{bhardwaj2026highrateqldpcprocessors}. The gadget introduces three sets of qubits. The \textit{vertex checks} $V$ are new $X$-type stabilisers, one per data qubit of $S$, each acting on its data qubit and on a few new ancilla qubits. Their product holds the value of $\bar X$. These new ancillas, the \textit{edge ancillas} $E$, are initialised in $\ket{0}$, kept for the whole measurement and finally measured in the $Z$ basis: they guarantee the randomness of the individual vertex checks, and play the role of data qubits for the gadget. Lastly, the \textit{cycle checks} $C$ are new $Z$-type stabilisers acting on edge ancillas only: they gauge out any additional degree of freedom of the deformed code. On the code side, $X$ checks are left unchanged, while the $Z$ checks overlapping $S$ are \textit{deformed}: they are extended onto edge ancillas so as to keep commuting with the vertex checks. The gadget may also be pictured as a graph whose vertices are the vertex checks and whose edges are the edge ancillas, each joining the vertex checks it interacts with; the cycle checks are then its independent cycles. Every edge ancilla must meet $V$ an even number of times such that the ancilla contributions cancel in the product of all vertex checks, rightfully returning $\bar X$.

In matrix form, with columns indexed by the $n$ data qubits and by the edge ancillas, the checks of the merged code read
\begin{equation} \label{eq:merged}
\begin{aligned}
    H_X^{\text{merged}} &= \bordermatrix{ & [n] & E \cr \text{original} & H_X & 0 \cr V & \Pi & \Gamma^{\top}}, \\[14pt]
    H_Z^{\text{merged}} &= \bordermatrix{ & [n] & E \cr \text{original} & H_Z & J \cr C & 0 & \Lambda}.
\end{aligned}
\end{equation}
Here $\Pi$ attaches each vertex check to its data qubit, with a single nonzero entry per row; $J$ deforms the $Z$ checks onto the edge ancillas; and $\Gamma^\top$ and $\Lambda$ are the respective $X$ and $Z$ parity check matrices of the gadget.
The merged checks commute if $\Pi H_Z^{\top}=\Gamma^{\top}J^{\top}$, \textit{i.e.} the deformation compensates the overlap of every $Z$ check with the support, and if $\Lambda\Gamma=0$, \textit{i.e.} every cycle check meets every vertex check an even number of times.

The simplest gadget, the \textit{skeleton}, is read off the code itself. There is one vertex check per qubit of $S$ and one edge ancilla per $Z$ check overlapping $S$, so up to relabelling:
\begin{align}
    \Pi \sim I \label{eq:pi_skeleton}\\
    J \sim I \label{eq:J_skeleton}
\end{align}
where $\sim$ indicates equality on relevant restrictions (respectively on $S$ and on the set of the $Z$ checks intersecting $S$). Then from the commutation relations above:
\begin{equation}
    \Gamma \sim H_Z \label{eq:Gamma_skeleton}
\end{equation}

The skeleton is a valid gadget, but in general not a fault-tolerant one. Indeed, the resulting deformed code may have a reduced code distance (with logical operators finding a shorter path through the gadget). A sufficient condition to prevent this is for the graph representing the gadget to be a good expander, with a Cheeger constant $h\geq1$~\cite{bhardwaj2026highrateqldpcprocessors, ide2025homological, williamson2026gauging}. Gadgets are therefore \textit{thickened}: edges are added until $h\geq1$. Each added edge creates new independent cycles, and a cycle check is added for every one of them, so that the gadget introduces no logical qubit of its own. In \cref{eq:merged}, thickening adds rows to $\Gamma$ and $\Lambda$ (and zero columns to $J$), while $V$ and $\Pi$ remain those of the skeleton: only $E$ and $C$ grow.

Mitten codes make this construction particularly economical~\cite{bhardwaj2026highrateqldpcprocessors}. Following Ref.~\cite{bhardwaj2026highrateqldpcprocessors}, we label their four left data blocks $D_1,\ldots,D_4$, their right data block $D_5$, and their check blocks $X_0$, $X_1$, $Z_0$ and $Z_1$. Their logical $X$ operators admit a canonical basis $\{\bar X_g\}_{g\in G}$ in which every operator is supported on $D_1\cup D_2$ only, and is obtained from $\bar X_e$ by translating the labels of its qubits, $q\mapsto gq$:
\begin{equation}
    \text{supp}(\bar X_g) = g~\text{supp}(\bar X_e)
\end{equation}
On these two blocks, only the $Z_0$ checks act, via the blocks $R(b_0)$ and $R(b_1)$ (see \cref{eq:parity_check_mitten}), which by \cref{eq:regular_reps} are invariant under a simultaneous left translation of all qubit and check labels. Since a surgery gadget is specified entirely by this information --- which data qubits carry its new $X$-checks, and which $Z$-checks must be deformed onto its ancillas --- its ancilla system and internal checks are identical for every $g$, and only the interface is relabelled. A single \textit{seed} gadget, built for $\bar X_e$, therefore measures any $\bar X_g$ once its interface is rewired by $g$. The same holds for $Z$-type measurements, whose canonical operators live on $D_1\cup D_3$ and are acted on by $X_0$.

\subsection{Logic in NeST}

\begin{figure}
    \centering
    \includegraphics[width=\linewidth]{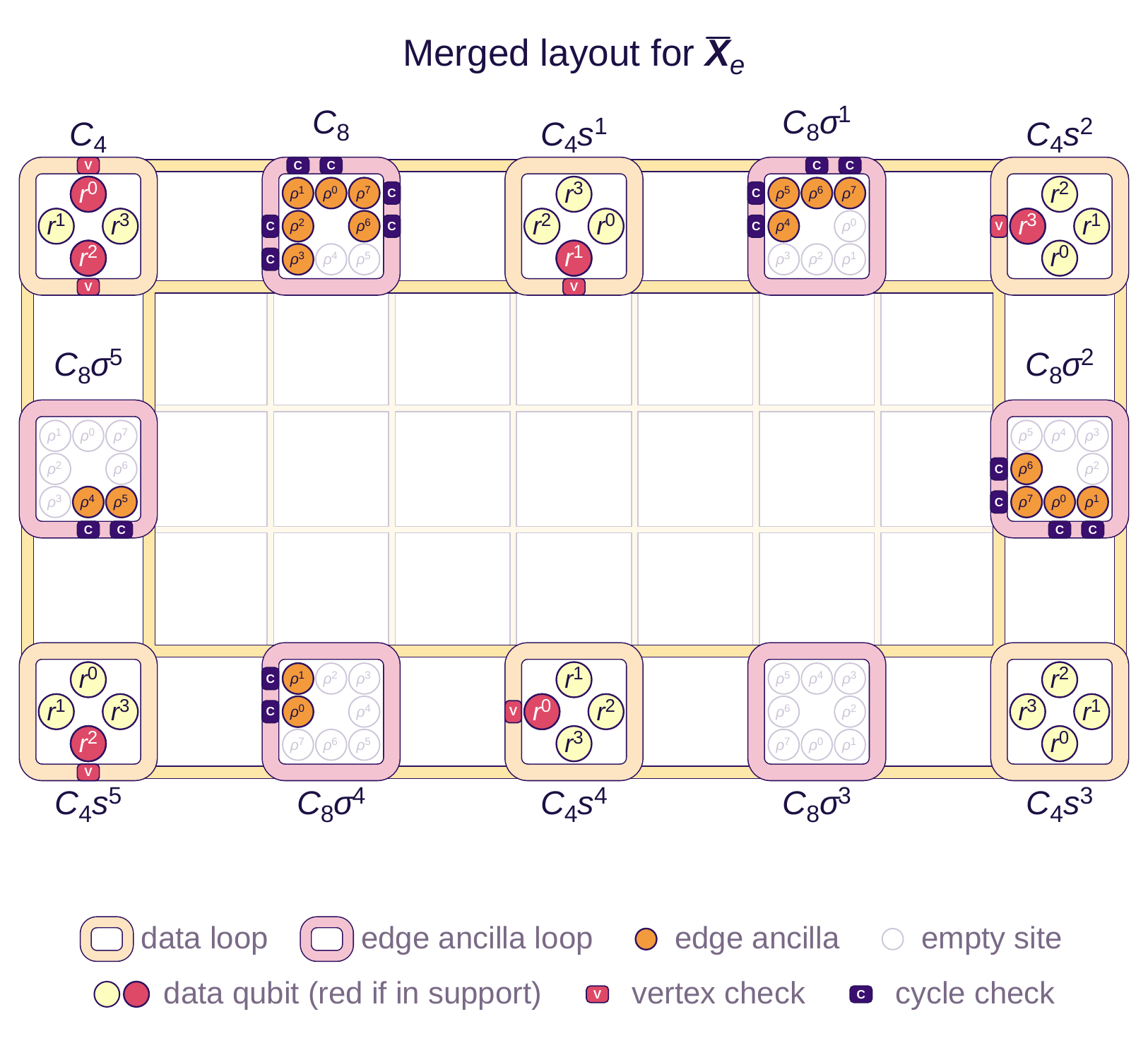}
    \caption{\textbf{Merged layout for the measurement of $\bar X_e$ on NeST}, zooming on one of the five unit cells (that of $D_1$). The layout is the same as in \cref{fig:group_embedding}, with additional qubits placed in previously empty loops. Data qubits occupy the same locations as before and those in the support of $\bar X_e$ are shown in red. Edge ancillas sit in between data loops and are indexed using the group $\tilde G=C_{mn_1}\rtimes C_{n_2}$, with generators $\rho$ and $\sigma$. Since the support of $\bar X_e$ is irregular, only some sites are populated (orange), while the others remain empty (white). All data and edge ancillas are static and all checks are shuttled. There is one vertex check per support data qubit, riding the belt of its data loop before moving to the edge ancilla loops. The cycle checks only ever interact with the edge ancillas. Note that in our scheduling $V$ and $C$ normally do not coexist on the lattice. Besides, the code checks $X_i$ and $Z_i$, which keep their memory schedule (up to deformation of the $Z$ checks), are omitted for clarity. The measurement of $\bar X_g$ is simply obtained by left-multiplying all data and edge ancillas by $g$.}
    \label{fig:surgery_embedding}
\end{figure}

We now map this construction onto NeST. In the spirit of \cref{sec:architecture}, the data qubits of the gadget --- its edge ancillas --- are kept static while all checks are shuttled. The resulting layout is shown in \cref{fig:surgery_embedding}. Data blocks $D_i$ remain as in memory, and so do the $X_i$ check blocks. The $Z_i$ blocks also keep their memory schedule, with a single addition: every deformed $Z_0$ check interacts with one extra edge ancilla qubit, which costs one CZ layer and a short excursion. Edge ancillas sit in what used to be transfer loops, \textit{i.e.} buffers serving to avoid collisions during inter-loop transfers. As we will see, a NeST-compatible gadget may require $|E|>|G|$ after thickening. Rather than splitting the edge ancillas across multiple unit cells, we rather group them all in the same cell as data qubits $D_1$. This reduces inter-cell shuttles. Their hosting loops may then hold more than $n_1$ sites, as long as the inter-qubit spacing in those loops, $\ell_{qq}=4l_e/(mn_1)$, stays greater than 200nm~\cite{desmet2024highfidelitysinglespinshuttlingsilicon} (remember $l_e=1\mu$m is the edge length). Two shuttled species are added: the vertex checks $V$ and the cycle checks $C$. As there is exactly one vertex check per support data qubit in $D_1\cup D_2$, we further split $V$ into two kinds $V_1$ and $V_2$, which ride the belts of the hosting loops like code checks, visiting first their data qubit and then their edge ancillas, across both unit cells of $D_1$ and $D_2$. The cycle checks $C$ ride the belts of the edge-ancilla loops and only ever interact with edge ancillas: they therefore always live in the cell of $D_1$. A merged stabiliser round runs the $X_i$, $Z_i$, $V_i$ and $C$ phases one after the other, each occupying the lattice in turn.

Recall that a specific feature of the mitten codes is the reusability of the gadget (last paragraph of the previous subsection). To exploit this on NeST, we label every new qubit with a group element, which fixes its site on the lattice. Each vertex check is attached to a single data qubit of the support, whose label it takes ($V$ does not grow during thickening). Skeleton edge ancillas take the label of the $Z_0$ check they come from. After thickening, however, edge ancillas may number more than $|G|$. We therefore subdivide $E$ sites into $m$ blocks indexed by $G$, so that the edge-ancilla loop following the data loop $C_{n_1}s^l$ holds $mn_1$ evenly spaced sites (\cref{fig:surgery_embedding} shows $m=2$). We may use two equivalent indexations to describe $E$ sites: either index every block independently using $G$, or encode the sites altogether via a larger group $\tilde G$ equal to $G$, except that its first factor $C_{n_1}$ in \cref{eq:G_n_sigma} is replaced with $C_{mn_1}$. Cycle checks, which interact with edge ancillas only, are labelled by $\tilde G$. This group is of the form $G(\vec{n},\sigma)$ (\cref{eq:G_n_sigma}), thus its group actions are rigid belt motions too. Since the support of the measured logical operator is irregular and does not cover the entire lattice, $E$, $V$ and $C$ sites are not all populated. We however stress that even empty sites carry a label (as seen in \cref{fig:surgery_embedding}) and that this is crucial to preserve the regularity of the group structure. Rewiring the gadget from $\bar X_e$ to $\bar X_g$ amounts to a simultaneous left translation of all labels. This changes which sites are occupied, but neither the lattice nor the motions, so that NeST can seamlessly reuse the gadgets as Ref.~\cite{bhardwaj2026highrateqldpcprocessors} does. In the following, we thus focus on a single logical measurement, and without loss of generality on $\bar X_e$. The challenge is now to make this specific measurement NeST-compatible.

With every qubit labelled, rigidity becomes a condition on the matrices of \cref{eq:merged}: qubits move as whole trains only if every block of $\Pi$, $\Gamma$, $J$ and $\Lambda$, coupling a block of checks to a block of qubits, is a group multiplication $L(\alpha)$ or $R(\alpha)$, with $\alpha$ in the group algebra of $G$ or $\tilde G$, exactly as in \cref{sec:architecture}. Each element of $\alpha$ is then one rigid motion of a train, \textit{i.e.} one gate layer. Such a block may be restricted to rows and columns of chosen sites --- CZ gates can be applied selectively. The skeleton satisfies this condition for free for $\Pi$, $J$ and $\Gamma$, given \cref{eq:pi_skeleton,eq:J_skeleton,eq:Gamma_skeleton} ($H_Z$ is by design NeST-compatible). Thickening is where rigidity becomes non-trivial. An unconstrained optimiser adds edges between arbitrary pairs of vertices. Each of them is still a restricted block $R(\alpha)$ on its own, but different edges generally call for different couplings $\alpha$, each with its own sites and motions. 

We thus follow a specific procedure for thickening. The first step is to grow $E$, which we allow for by doubling the number of edge ancilla sites ($m=2$). These are split between two cosets $E_{\text{skel}}$ and $E'$, hosting skeleton and post-thickening edge ancillas respectively. At this step a new $\Gamma$ is computed ($J$ is only expanded with zero columns and $\Lambda$ is addressed at the next step). The goal of the routine is to populate the new sites in the most frugal way that would satisfy all the gadget requirements. On the hardware-agnostic side, it minimises the size of $E$ subject to the constraints that $(i)$ the Cheeger constant satisfies $h\geq1$ (no distance reduction) and $\Gamma$ has $(ii)$ even-parity rows (correct measurement of the logical) and $(iii)$ low-weight columns (LDPC). NeST further imposes that the overall block is compatible with the conveyor-belt shuttling \textit{i.e.}
\begin{equation} \label{eq:Gamma_nest_condition}
    \Gamma^\top \sim \bordermatrix{ & E_{\text{skel}} & E' \cr V_1 & R(b_0^*) & R(\alpha_1) \cr V_2 & R(b_1^*) & R(\alpha_2)}
\end{equation}
for a certain $\alpha=(\alpha_1,\alpha_2) \in \mathbb{F}_2[G]^2$ to be found. This is solved via integer programming as explained in \cref{app:surgery}.

The second step builds $C$ and $\Lambda$. The procedure is similar, with general hardware-agnostic requirements \textit{e.g.} $\Lambda\Gamma=0$, as well as a NeST-specific constraint, namely:
\begin{equation} \label{eq:Lambda_nest_condition}
    \Lambda \sim R(\gamma^*), \qquad \gamma\in\mathbb{F}_2[\tilde G].
\end{equation}
Note how $\gamma$ here lives in $\mathbb{F}_2[\tilde G]$ rather than $\mathbb{F}_2[G]$: this is because both $C$ and $E$ are indexed by such elements. $C$ and $\gamma$ are found by a greedy search that assembles the cycle checks from translations of the short cycles of the gadget graph while keeping $|\gamma|$ small, as detailed in \cref{app:surgery}.

\cref{tab:gadgets} compares the resulting gadget with that of Ref.~\cite{bhardwaj2026highrateqldpcprocessors} for the $\llbracket 150,30,10\rrbracket$ code, built upon $C_3\rtimes C_{10}$ ($|G|=30$). Rigidity costs no qubits at all: both gadgets have $34$ edge ancillas, $18$ vertex checks and $17$ cycle checks, the same largest check weight and a Cheeger constant of exactly one. We verified with an integer program that the merged code has distance exactly $10$, like the original code. We also simulated the full merge-measure-split protocol for all $30$ rewirings $\bar X_g$: the gadget measures exactly $\bar X_g$ while preserving the $29$ other logical qubits (\cref{app:surgery}).

The price of rigidity is paid in time instead. Without constraints, cycle checks can all be measured in six gate layers, and so can the vertex checks. On NeST, they must follow rigid motions, such that the $17$ cycle checks take $12$ gate layers and the vertex checks $9$, resulting in a merged round containing $40$ CZ layers against at least $31$ in Ref.~\cite{bhardwaj2026highrateqldpcprocessors}. Routing these motions to estimate the final syndrome extraction time and simulating the merged circuit are left for future work, as is the extension to Pauli-product and parallel measurements. These results nevertheless establish the compatibility of mitten-code logic with NeST, as we were able to build a measurement gadget whose every qubit moves by rigid, globally controlled belt motions, with as many qubits as the unconstrained one.

\begin{table}[htbp]
  \centering
  \begin{tabular}{lcc}
    \toprule
     & Ref.~\cite{bhardwaj2026highrateqldpcprocessors} & This work \\
    \midrule
    Rigid belt motions only & no & yes \\
    Edge ancillas $E$ & $34$ & $34$ \\
    Vertex checks $V$ & $18$ & $18$ \\
    Cycle checks $C$ & $17$ & $17$ \\
    Maximal check weight & $10$ & $10$ \\
    Cheeger constant $h$ & $1$ & $1$ \\
    Merged-code distance & $10$ & $10$ \\
    CZ layers, $V$ / $C$ phases & $6$ / $6$ & $9$ / $12$ \\
    CZ layers per merged round & $31$ & $40$ \\
    \bottomrule
  \end{tabular}
  \caption{\textbf{Gadgets measuring $\bar X_e$ on the $\llbracket 150,30,10\rrbracket$ code}: the unconstrained gadget of Ref.~\cite{bhardwaj2026highrateqldpcprocessors} against our NeST-compatible one. Counts are the qubits the measurement uses; rewiring the gadget to another $\bar X_g$ uses as many, on translated sites. The maximal check weight includes the deformed $Z_0$ checks, and the merged-code distances are exact, obtained by integer programming. Gate layers are counted in the unit cell of $D_1$, which hosts the edge ancillas. The gadget of Ref.~\cite{bhardwaj2026highrateqldpcprocessors} does not move rigidly, so we assign it the minimum any schedule needs, set by the largest number of interactions of a single check or data qubit: six in both phases. Our gadget has the same minimum, so that rigidity costs $9$ additional CZ layers per merged round.}
  \label{tab:gadgets}
\end{table}

\section{Discussion} \label{sec:discussion}

In this paper, we have shown that a silicon spin qubit processor can run a $20\%$-rate qLDPC code with a syndrome-extraction cycle of a few tens of microseconds, between two and three orders of magnitude faster than the reconfigurable-atom implementations these codes were designed for. Thresholds are above $10^{-3}$ under every shuttling-noise assumption we considered. Extrapolating the simulations, our $\llbracket 1400,280,\leq 28\rrbracket$ code would reach a per-logical-qubit per-round error under $10^{-14}$ assuming a ten-fold improvement over early shuttling demonstrations with other physical errors at $p=0.1\%$, with only $9$ physical qubits (data and ancilla) per logical qubit. The surface code would require two orders of magnitude more qubits in this regime. We do not believe this shuttling requirement to be so stringent in a mature device, as spin shuttling experiments are quickly evolving, with theoretical works showing that there should still be room for improvement \cite{binder_shuttling_2025, Langrock_2023}. Beyond memory, we have also made a step towards logic by constructing efficient graph surgery gadgets tailored to our spin-qubit processor.

This performance rests on a co-design rather than on a mapping. NeST, the architecture behind these numbers, is essentially a plain square lattice of shuttling tracks, where the pathways are interpreted as a hierarchy of nested conveyor-belt loops. It was arrived at by asking what the hardware does naturally rather than how to force a code onto it. A belt advances every qubit it carries by the same number of sites, in either direction, under a single global control signal and regardless of how many qubits ride it. That is a cyclic shift, so one loop reproduces multiplications in a cyclic group cheaply. Arranging multiple loops in a larger ring and passing qubits between them repeats the construction one tier up. The non-abelian part costs nothing extra: a train of ancillas can enter its destination loop in either direction, and the two choices leave it in opposite orders, so the inversion $r\mapsto r^{-1}$ is obtained simply by choosing an approach direction. The groups generated by these two motions alone are the iterated semidirect products of cyclic groups, and NeST mimics their geometry exactly --- every left and right multiplication becomes a rigid, globally controlled motion of whole trains, no qubit is ever addressed individually, and data qubits need not move. This is also why a direct port of the neutral-atom construction would not have worked. There, any permutation is available because the atoms move through free space; here the lattice fixes which permutations exist, and two ancillas meeting is a forbidden collision. Starting from the code would have left us routing arbitrary group elements through a track network --- the generality that code-agnostic schedulers \cite{chadwick2026cablecarefficientlyschedulingqldpc} pay for in routing depth. Starting from the hardware instead yields a group family whose every element is a single rigid motion, which is what keeps the shuttling tolerable. The family is broad enough that the restriction costs little: it contains every finite abelian group, hence every quasi-cyclic lifted product, and a search of its members returned rate-$20\%$ codes with parameters comparable to the best published at each distance, and a candidate distance-28 code.

Several directions follow. The most immediate regards the optimisation of some of our scheduling routines. In particular, for simplicity, our workflow first fixed the gate ordering before routing the motions. The two interact strongly however, and jointly optimising the group representation $(\vec{n},\sigma)$, the gate ordering (potentially interleaving $X$ and $Z$ stabilisers) and the routes together should cut the shuttling distance further --- as should relaxing the parity convention of \cref{eq:ni_even}, which was adopted only for layout simplicity. Another idea would be to overlap all five unit cells composing the simulated mitten codes, by interleaving different-cell qubits in the same loop (in the same fashion as Ref.~\cite{cai_looped_pipelines_2023}). This would essentially eliminate the inter-cell shuttling terms.

A second concerns logic. We have shown that a NeST-compatible gadget exists for the measurement of $\bar X_g$ on one of our codes, and verified the protocol without noise. What remains is to route its motions on the lattice, cost the merged cycle in microseconds and run circuit-level simulations. Beyond that, measuring Pauli products requires gadgets spanning two blocks, and running measurements in parallel requires several gadgets to share the lattice without colliding. Whether the thickening procedure of \cref{sec:logic} generalises to the larger codes of \cref{tab:target_codes}, where the integer programs grow quickly, is also open, as is the question of whether the native group actions could supply automorphism-based gates more cheaply than surgery.

A third one is the group family itself. Relaxing $srs^{-1}=r^{-1}$ to $srs^{-1}=r^{a}$ now asks for neighbouring loops to run at different speeds --- this is the generalisation of the mode A and B loops --- which should not represent a massive experimental leap. It however also replaces reflections with more elaborate permutations, which junctions can supply at some scheduling cost (see Fig. 17 of \cite{siegel2025snakesplanemobilelow}). Mapping that trade-off would widen the implementable set considerably. 

Fourth, we focused on mitten codes for their simple protograph, but \cref{fig:main_figure} accommodates any protograph shape, and wider or taller base matrices trade cell count against check weight in ways worth exploring.

Finally, two noise questions are specific to this platform. Shuttling and idling noise are strongly $Z$-biased. We designed our stabiliser circuits such that the former would exclusively land on ancillas --- data qubits are kept static and interact with ancillas via CZs --- and the latter is suppressed via Hahn-echo. If one of these assumptions broke down, the noise injected on data qubits could show a strong $Z$ bias, motivating the search for biased codes \cite{bonilla_ataides_xzzx_2021, roffe_bias_tailored_2023, tuckett_tailoring_2019} or more simply the use of syndrome extraction protocols exploiting correlations (\textit{e.g.} measuring $X$ and $Y$ stabilisers rather than $X$ and $Z$). Besides, two-level fluctuators \cite{paladino_1f_2014, connors_low_2019, jnane2026electronshuttlingprobecharge} in the form of charge defects can spontaneously appear in a solid-state chip and coherently corrupt all qubits shuttled in their vicinity.
This could create a correlated failure mode that no amount of code distance addresses. Concatenation with a small outer code may be the right answer here \cite{gidney_yoked_2025}.

\section{Acknowledgements}
I am very grateful to Simon Benjamin, Armands Strikis and Tim Chan for useful conversations throughout. I acknowledge the use of Claude AI for the numerical simulations, figures and parts of the text.

\bibliography{ref}

\newpage
\appendix
\crefalias{section}{appendix}
\crefalias{subsection}{subappendix}
\onecolumngrid

\section{Shuttling distance upper bound in the two-tier embedding} \label{app:2_tier_shuttling_upper}

We here estimate the per-ancilla cost of the motions of the two-tier embedding in terms of shuttling distance. Let us consider again the left or right multiplication by $r^is^j$, which translates into an inter-loop hop of $j$ loops plus an intra-loop rotation of $i$ increments, potentially coupled with a reflection. The first step (panel (c)) of a inter-loop transfer always costs 3 edges (whether it involves a corner loop or not). Every subsequent loop-to-loop transfer is 4 edges. The final step consisting of wrapping around the destination loop (panels (d) and (e)) costs 4 edges as well, whether a reflection is required or not. The movement is then completed by a final intra-loop rotation of $L_{\text{intra}}=i+\delta i$ inter-qubit increments, where $\delta i$ is the index shift the qubits end up with after a global rotation ($\delta i=0$ for a reflection). By remembering that the loop size is related to the inter-qubit spacing $\ell_{qq}$, such that one edge length is $\ell_e = n_1\ell_{qq}/4$, one obtains a shuttling distance:
\begin{equation*}
    L \leq (3 + 4j) \times n_1\ell_{qq}/4 + L_{\text{intra}}\ell_{qq}.
\end{equation*}
$L_{\text{intra}}$ and $j$ are respectively bounded by $n_1/2$ and $n_2/2$ (if they exceed these values one can rotate in the opposite direction). This leads to the final bound:
\begin{equation} \label{eq:2_tier_shuttling_upper}
L \leq \frac{\ell_{qq}\,n_1}{4}\,(2n_2 + 5)
\end{equation}
The dominating term of the shuttling distance is $n_1n_2\ell_{qq}/2=|G|\,\ell_{qq}/2$. This is expected and corresponds to qubits having to travel from the group $C_{n_1}s^l$ to $C_{n_1}s^{n_2/2+l}$ \textit{i.e.} across two opposite corners of the lattice. Importantly, this describes a no-collision fully-synchronised schedule of the ancillas. While this only represents one unit cell of the code, we can note that the upper bound we found is roughly $|G|/2$ inter-qubit spacings. This is half the linear extent of a lattice holding $|G|$ qubits, thus satisfying the shuttling distance criterion given at the start of \cref{sec:architecture}. Most motions may also involve shorter transfers than what this upper bound prescribes. The presence of the optional shuttling tracks of \cref{fig:group_embedding} may further reduce the shuttling distance as well, when using a group-agnostic automated route planner.

\section{Compact embedding} \label{app:compact_embedding}

\begin{figure}
    \centering
    \includegraphics[width=.6\linewidth]{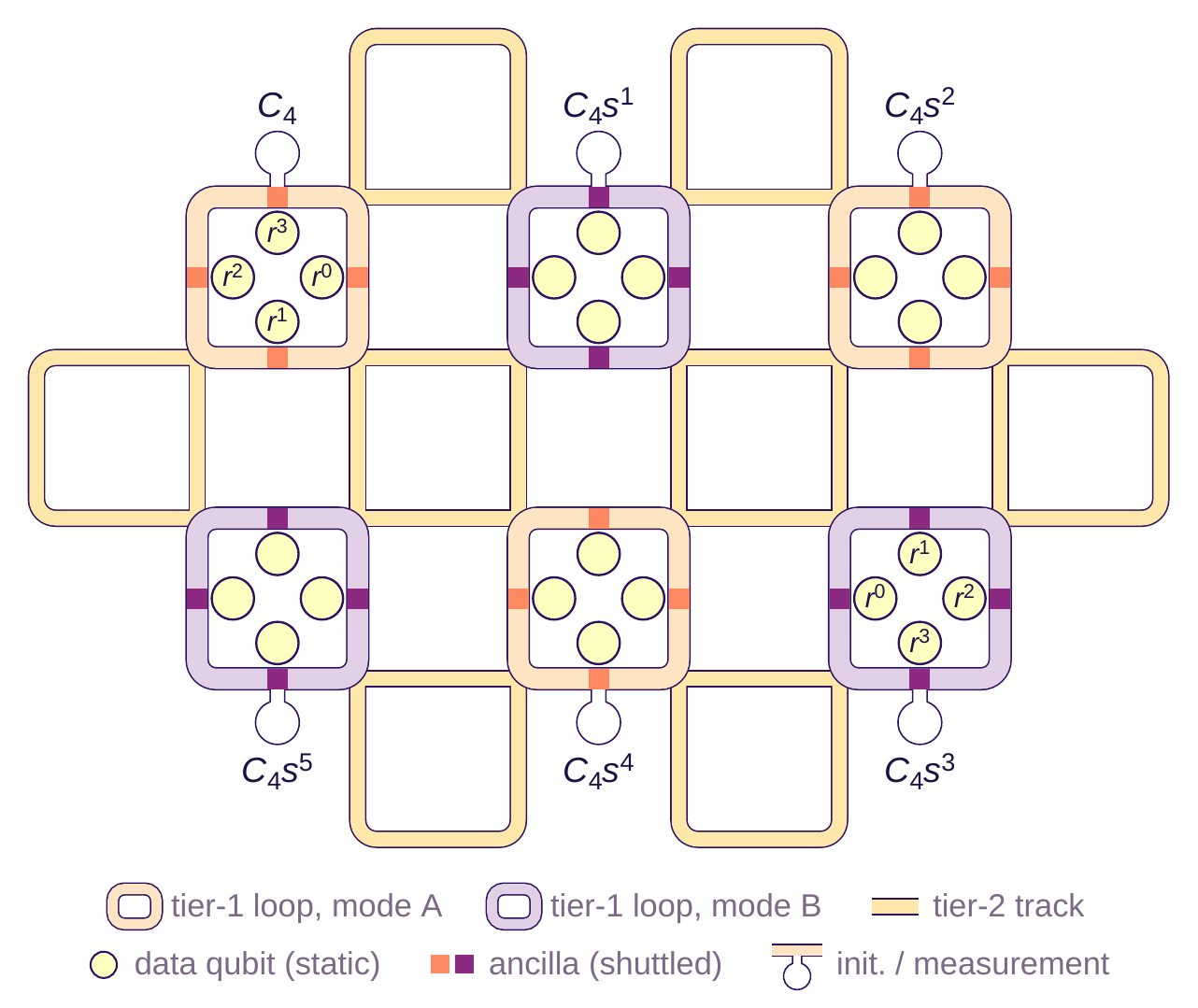}
    \caption{\textbf{Compact two-tier embedding for $n_1=4$ and $n_2=6$.} The six groups $C_4 s^{l}$ ($l = 0,\dots,5$) are hosted on tier-1 loops arranged in a $3 \times 2$ array, like in \cref{fig:group_embedding}. The qubits indexations are also identical. The only difference with the main text embedding resides in the spacing between tier-1 loops, reduced to a single edge. Tier-2 motions may involve the protrusions to avoid collisions.}
    \label{fig:group_embedding_compact}
\end{figure}

We give in \cref{fig:group_embedding_compact} an alternative layout for a two-tier unit cell. Compared to \cref{fig:group_embedding}, this one is about four times more compact in terms of area as there is here only one edge (instead of three) between neighbouring tier-1 loops. The main text stated that a three-edge spacing was necessary to avoid collisions: as qubits uncurl from their departure loop, consecutive groups follow each other in lockstep with no additional space (see panel (c) of \cref{fig:motions}). This is however an artifact of qubits travelling in a mostly straight line. This compact layout recovers this feature by having qubits zigzag through the protrusions instead of taking the shortest route.

\section{Details on the code search} \label{app:code_search}

In order to find high-distance codes based on the target group family $G(\vec n,\sigma)$, we initiated a deep code search with the help of the tool sQetch~\cite{sqetch}, a GPU-accelerated random information-set distance estimator introduced in Appendix~H of Ref.~\cite{bhardwaj2026highrateqldpcprocessors}. The search was parametrised by the code size $n$ and the target distance $d$. For $n = 5\ell \in [300,700]$ ($\ell=|G|$), there were 291 non-abelian groups of the form $G(\vec n,\sigma)$ with $\ell = 60$–$140$ to scan (up to isomorphism). We kept the mitten shape, $(1\times2,1\times2)$ with weight-3 entries as defined in \cref{eq:parity_check_mitten}. In particular, one entry of both $A$ and $B$ contains the identity, $a_1 = e + g_4 + g_5$ and $b_1 = e + h_4 + h_5$, and is chosen so that $L(a_1)$ and $R(b_1)$ are full rank (required for logic). This results in $k=\ell$ thus a guaranteed rate $k/n = 1/5$. Theorem~2 of Ref.~\cite{bhardwaj2026highrateqldpcprocessors} upper-bounds $d_X$ and $d_Z$ by the distances of the two base codes, so we screened classical base codes first and only then combined them into quantum codes. A census of 200 random base codes per group, with distances estimated by sQetch, ranked the groups by how often they yielded high distances. For the highest-ranked groups we then filled two pools, one for $A$ and one for $B$, both clearing the target distance. Translating the entries of $A$ or $B$ by group elements gives an equivalent code, so each pool kept only one base code from every class of translates. As a short sQetch run can miss a base code's lightest codewords thus overestimate the distance, every pooled base code was re-measured with $2\times10^{7}$ trials and dropped if it fell below the target. Each pair of base codes, which defines a quantum code, was discarded as soon as it failed one of three tests, run cheapest first: its Tanner graph must be connected, an exhaustive search must find no logical of weight $\leq 4$, and a quick sQetch run of $2\times10^{5}$ trials each for $d_X$ and $d_Z$ must find no logical below the target. The pairs that passed were verified more thoroughly. First, the distances of the two base codes, which cap $d_X$ and $d_Z$, were re-measured with sQetch and, for the codes up to $n=700$, checked with the \texttt{ldpc} package. Second, sQetch ran $10^{8}$ trials each for $d_X$ and $d_Z$ on the full code, over five seeds. Third, an exhaustive search ruled out logicals of weight $\leq 5$. sQetch only finds codewords, so each distance it reports is an upper bound, backed by an explicit logical of that weight. The exhaustive search supplies the proven lower bound $d \geq 6$. The $\llbracket 1400,280,\leq 28\rrbracket$ code comes from a second search at $n=5\ell\in[1300,1560]$, restricted to the 62 groups $G(\vec n,\sigma)$ with $\ell=260$–$312$ whose largest abelian subgroup has index at least 4 (in groups of index 2 we found no base code above distance 12). It followed the same procedure with a census of 100 base codes per group and a screening run of $10^{6}$ trials each for $d_X$ and $d_Z$.
At this size the null-space basis of the full code no longer fits sQetch's on-chip GPU memory, so each sQetch call ran on a shortened code: a fresh random set of columns is deleted and only codewords vanishing there are searched. Every reported weight is therefore still that of a logical of the full code. We validated this mode by reproducing the distances of the eight codes of \cite{bhardwaj2026highrateqldpcprocessors} and of our three smaller codes.
The \texttt{ldpc} check of the base codes is not reliable at this size, so the base-code distances of the $\llbracket 1400,280,\leq 28\rrbracket$ code rest on sQetch alone.

\section{Full description of the target codes} \label{app:codes}

\cref{tab:target_codes_full} gives a full description of the target codes presented in the main text.

\begin{table*}[tbp]
  \centering
  \small
  \setlength{\tabcolsep}{5pt}
  \begin{tabular}{@{}l l l c c@{}}
    \toprule
    $\llbracket n,k,d \rrbracket$ & Group and generators & Supports $A=[a_0\,\|\,a_1]$, $B=[b_0\,\|\,b_1]$ & $d_{\mathrm{circ}}$ & SE time \\
    \midrule
    $\llbracket 150,30,10 \rrbracket$ & \begin{tabular}[t]{@{}l@{}}$C_{3} \rtimes C_{10}$ \\ $r^{3}=s^{10}=e$ \\ $s$ inverts $r$\end{tabular} & \begin{tabular}[t]{@{}l@{}}$a_0=\{e,\, r s^{7},\, r^{2} s^{6}\}$ \\ $a_1=\{e,\, r s^{5},\, s^{6}\}$ \\ $b_0=\{r s^{4},\, r s^{6},\, r^{2} s^{2}\}$ \\ $b_1=\{e,\, r s^{9},\, s^{6}\}$\end{tabular} & $\leq 8$ & $16.4~\mu$s \\
    \addlinespace
    $\llbracket 200,40,12 \rrbracket$ & \begin{tabular}[t]{@{}l@{}}$C_{5} \times C_{4} \rtimes C_{2}$ \\ $r^{5}=s^{4}=t^{2}=e$ \\ $s$ fixes $r$ \\ $t$ inverts $r$; fixes $s$\end{tabular} & \begin{tabular}[t]{@{}l@{}}$a_0=\{r s^{2},\, r^{2} s t,\, r^{3} s t\}$ \\ $a_1=\{e,\, r^{2} s,\, r^{2} s^{2}\}$ \\ $b_0=\{r^{4},\, r^{4} s^{3},\, s\}$ \\ $b_1=\{e,\, r^{3},\, r^{3} s t\}$\end{tabular} & $\leq 10$ & $17.5~\mu$s \\
    \addlinespace
    $\llbracket 300,60,14 \rrbracket$ & \begin{tabular}[t]{@{}l@{}}$C_{6} \rtimes C_{10}$ \\ $r^{6}=s^{10}=e$ \\ $s$ inverts $r$\end{tabular} & \begin{tabular}[t]{@{}l@{}}$a_0=\{r^{2} s^{3},\, r^{3} s^{2},\, r^{4} s^{4}\}$ \\ $a_1=\{e,\, r s^{6},\, r^{3} s^{8}\}$ \\ $b_0=\{r,\, s^{4},\, s^{7}\}$ \\ $b_1=\{e,\, r^{2} s^{6},\, r^{5} s^{2}\}$\end{tabular} & $\leq 13$ & $20.7~\mu$s \\
    \addlinespace
    $\llbracket 480,96,\leq 16 \rrbracket$ & \begin{tabular}[t]{@{}l@{}}$C_{3} \times C_{4} \rtimes C_{4} \rtimes C_{2}$ \\ $r^{3}=s^{4}=t^{4}=u^{2}=e$ \\ $s$ fixes $r$ \\ $t$ inverts $s$; fixes $r$ \\ $u$ inverts $r$, $s$; fixes $t$\end{tabular} & \begin{tabular}[t]{@{}l@{}}$a_0=\{r s t^{2},\, r s^{3} t u,\, r^{2} s^{2} t^{2}\}$ \\ $a_1=\{e,\, r s t^{3},\, s^{3} t\}$ \\ $b_0=\{e,\, r^{2} s^{2} t,\, t\}$ \\ $b_1=\{e,\, r s t^{2},\, r^{2} s t^{3} u\}$\end{tabular} & $\leq 15$ & $26.1~\mu$s \\
    \addlinespace
    $\llbracket 600,120,\leq 18 \rrbracket$ & \begin{tabular}[t]{@{}l@{}}$C_{5} \rtimes C_{6} \rtimes C_{4}$ \\ $r^{5}=s^{6}=t^{4}=e$ \\ $s$ inverts $r$ \\ $t$ inverts $s$; fixes $r$\end{tabular} & \begin{tabular}[t]{@{}l@{}}$a_0=\{r^{3} s^{4} t^{3},\, s,\, t\}$ \\ $a_1=\{e,\, r s^{2} t^{3},\, r^{4}\}$ \\ $b_0=\{r s^{3},\, r^{3} s^{2} t^{3},\, r^{4} s^{3} t\}$ \\ $b_1=\{e,\, r s^{4} t^{2},\, r^{4}\}$\end{tabular} & $\leq 17$ & $29.3~\mu$s \\
    \addlinespace
    $\llbracket 700,140,\leq 20 \rrbracket$ & \begin{tabular}[t]{@{}l@{}}$C_{7} \rtimes C_{10} \rtimes C_{2}$ \\ $r^{7}=s^{10}=t^{2}=e$ \\ $s$ inverts $r$ \\ $t$ inverts $s$; fixes $r$\end{tabular} & \begin{tabular}[t]{@{}l@{}}$a_0=\{e,\, r^{3} s^{3},\, r^{5} s^{9}\}$ \\ $a_1=\{r^{4} s^{4},\, r^{5} t,\, s^{6}\}$ \\ $b_0=\{e,\, r s,\, r s^{8}\}$ \\ $b_1=\{r^{2} s t,\, r^{3} s^{9},\, r^{5} s^{3}\}$\end{tabular} & $\leq 20$ & $30.1~\mu$s \\
    \addlinespace
    $\llbracket 1400,280,\leq 28 \rrbracket$ & \begin{tabular}[t]{@{}l@{}}$C_{7} \rtimes C_{10} \rtimes C_{4}$ \\ $r^{7}=s^{10}=t^{4}=e$ \\ $s$ inverts $r$ \\ $t$ inverts $s$; fixes $r$\end{tabular} & \begin{tabular}[t]{@{}l@{}}$a_0=\{e,\, r^{2} s^{3} t^{2},\, r^{4} s^{3} t\}$ \\ $a_1=\{r^{4} s,\, s^{3} t^{2},\, t^{2}\}$ \\ $b_0=\{e,\, r s^{7},\, r^{5} t\}$ \\ $b_1=\{r^{5} s^{9} t^{2},\, r^{6} s^{3} t^{3},\, s^{9} t^{3}\}$\end{tabular} & --- & $\approx 42~\mu$s$^{\dagger}$ \\

    \bottomrule
  \end{tabular}
  \caption{
    Full description of the target mitten codes of \cref{tab:target_codes}, all at the $(1\times2,1\times2)$ shape with weight-3 entries (check weight $9$). Each group $G(\vec n,\sigma)$ is written as a chain of its cyclic tiers $C_{n_1}, C_{n_2}, \dots$, read left to right, each factor acting on all factors before it: $\times$ if it fixes all of them, $\rtimes$ if it inverts at least one. Below each group are its tier generators $r,s,t,u$ (tiers $1,2,3,4$) and their action on the lower tiers. Group elements are written lowest tier first, $r^i s^j t^k u^l$. Supports follow the lifted-product convention of \cite{bhardwaj2026highrateqldpcprocessors}, written out in \cref{eq:parity_check_mitten}. The first three codes are taken from \cite{bhardwaj2026highrateqldpcprocessors}, rewritten in the form $G(\vec n,\sigma)$; their distances are exact. The last four are obtained from the code search described in \cref{app:code_search}. $d_{\mathrm{circ}}$ and the syndrome extraction times are as in \cref{tab:target_codes}.
  }
  \label{tab:target_codes_full}
\end{table*}

\section{Details on the gate ordering optimiser} \label{app:gate_ordering}

To find good CZ gate orderings, we search over the sequence in which each ancilla block ($X_0$, $X_1$, $Z_0$, $Z_1$) performs its nine CZ gates. We restrict the search to orders with exactly two inter-cell switches, where the three interactions in each unit cell are performed consecutively (1296 orders per block), or exactly three, where one cell is split once (7776 orders per block). The figure of merit is the circuit distance $d_\mathrm{circ}$, the smallest number of faults that produces an undetectable logical error. In our circuits the only faults that spread to several data qubits are hook errors: an $X$ or $Y$ fault on an ancilla between two of its CZ gates propagates to every data qubit the ancilla interacts with afterwards. $d_\mathrm{circ}$ is therefore the minimum weight of a logical operator of an augmented code with one column per data qubit plus one per possible hook, which we estimate with a randomised information-set decoder \cite{prange1962,lee1988,leon1988,pryadko2022qdistrnd}. The search proceeds in stages of increasing cost. First, every block ordering is scored on its own. Next, for each check type, we combine the best orderings of its two ancilla blocks ($X_0$ and $X_1$, or $Z_0$ and $Z_1$) since they operate simultaneously, into a common sequence of CZ layers. This step aims at finding schedules with the fewest layers and idling, taking into account the fact that two ancilla blocks cannot occupy the same cell simultaneously. Since hook errors from both blocks can combine into a logical error, $d_\mathrm{circ}$ is then evaluated for each combined schedule as a whole. Finally, the most promising schedules are re-evaluated with $10^5$--$10^6$ information sets. Every low-weight fault pattern found is injected into a noiseless Stim circuit, where it must flip a logical observable without triggering any detector. Each reported $d_\mathrm{circ}$ is therefore a certified upper bound, while the absence of lighter patterns is a statistical statement. We select the ordering with the highest $d_\mathrm{circ}$, preferring fewer transfers at equal distance. Remaining ties are settled by short logical-error-rate simulations, since $d_\mathrm{circ}$ only certifies the lightest failing fault pattern, not how many such patterns there are.

\section{Hahn echo: pulse placement and noisy CZ gates} \label{app:hahn_echo}

In the main text, idling dephases a data qubit $q$ through the coherent rotation $R_z(\delta_q t)$ set by its quasi-static detuning $\delta_q$. Since $R_z$ commutes with CZ, all idle periods between two Hadamards on $q$ add up to a single rotation $R_z(\delta_q W_q)$, which one $X$ pulse at the midpoint of the idle time cancels through $R_z(\theta)XR_z(\theta)=X$, the leftover $X$ being absorbed into the Pauli frame. Here $W_q$ excludes the time spent in the qubit's own CZ gates, whose dephasing is part of the gate error. If the midpoint $W_q/2$ lands close to a CZ, a finite pulse may need to be slightly shifted: for $500$\,ns pulses, this leaves on average $0.18\,\mu$s of idle time unrefocused (at most $0.8\,\mu$s), a residual error of $1.3\times10^{-4}$ per window (at most $8\times10^{-4}$) by \cref{eq:idle}, which we neglect. 

Away from its calibrated working point, the exchange CZ is not diagonal: its flip-flop rotation in $\{\ket{01},\ket{10}\}$ is incomplete, adding an $XX$/$YY$ error. Pushing the idle rotations $D_k$ through the noisy gates $G_k=\mathrm{CZ}\,E_k$ is nevertheless exact,
\begin{equation}
    D_n G_n \cdots G_1 D_0 = D_\text{tot} \prod_{k=n}^{1} \mathrm{CZ}\,\bigl[D_{<k}^\dagger E_k D_{<k}\bigr],
\end{equation}
with $D_{<k}$ the rotation accumulated before gate $k$. Moving the idle rotations to the end of the window thus leaves two things. First, their sum $D_\text{tot}$ appears, which the echo pulse cancels exactly as for ideal CZs, since moving a rotation past the pulse inverts its sign. Second, each gate error $E_k$ is replaced by $D_{<k}^\dagger E_k D_{<k}$: the same error seen in a frame rotated about $Z$ by the phases $\phi_c$ and $\phi_t$ its two qubits accumulated before the gate. Such a change of frame leaves the error probability unchanged, and a depolarising channel, as simulated here, exactly unchanged. For a coherent exchange error (see \cref{app:spin_qubit_noise_model}), the $Z$-type part is unaffected, while the flip-flop amplitude $\ket{01}\leftrightarrow\ket{10}$ acquires the relative phase $\theta=\phi_c-\phi_t$, turning a fraction $\sin^2\theta$ of its $XX$/$YY$ weight into $XY$/$YX$. Averaged over the quasi-static detunings this fraction is $\tfrac12\bigl(1-\mathrm{e}^{-2\,\mathrm{Var}\,\theta}\bigr)$: $0.2$--$0.4\%$ of the CZ error in our schedules, and at most $1.3\%$, for a CZ infidelity of $10^{-3}$.

\section{Noise model specific to silicon spin qubits and performance evaluation} \label{app:spin_qubit_noise_model}

\subsection{Noise model}

While depolarising noise is the right first approach as it is general and provides insights independently of the hardware implementation details, it is a poor stand-in for the error processes that actually limit silicon spin qubits. Every QEC simulation reported below uses a noise model constructed from the physics of the platform and of our architecture in particular.

\subsubsection{Hamiltonians}

The Hamiltonian for a single qubit in the presence of a fixed magnetic field $\vec{B}_0 = (0, 0, B_0)$ is given in the $\{\ket{\uparrow},\ket{\downarrow}\}$ basis by
\begin{align}
    H = \mu_B g \vec{B}_0\cdot\vec{S} 
   = \frac{1}{2} \begin{pmatrix}
        \omega_0 & 0 \\
        0 & -\omega_0
    \end{pmatrix} 
    \label{eq:zeeman_split_hamil}
\end{align}
where $\hbar = 1$, $\omega_0 \coloneqq \mu_B g B_0$ is the Zeeman splitting (which we denote $E_z$), $\mu_B$ is the Bohr magneton, and $g\approx2$ the electron's $g$-factor. To perform single qubit gates, a transverse drive $\vec{B}_1 = (B_1\cos(\omega t + \phi), 0, 0)$ is added. Transforming to a frame rotating at the drive frequency $\omega$ and applying the rotating-wave approximation (RWA) yields,
\begin{align}
    H_\text{1Q} = \frac{1}{2}\begin{pmatrix}
        \delta \omega & \Omega e^{-i \phi} \\
        \Omega e^{i\phi} & -\delta \omega
    \end{pmatrix}
\end{align}
where $\delta \omega = \omega_0 - \omega$ is the detuning and $\Omega \coloneqq  \tfrac{1}{2}\mu_B g B_1$ is the Rabi frequency. Driving on resonance ($\delta\omega = 0$) enables universal single-qubit rotations in the $x-y$ plane by modulating the drive phase $\phi$.

Two-qubit gate are governed by the Heisenberg Hamiltonian \cite{meunier_cz, Burkard_cz}:
\begin{equation}
    H_\text{2Q} = \frac{1}{2}\begin{pmatrix} 2E_z & 0 & 0 & 0 \\ 0 & \Delta E_z - J & J & 0 \\ 0 & J & -\Delta E_z - J & 0 \\ 0 & 0 & 0 & -2E_z \end{pmatrix},
\end{equation}
where $J$ is the exchange coupling, $E_z = (\omega_1+\omega_2)/2$ and $\Delta E_z = \omega_1-\omega_2$. This system naturally yields a CZ for $J=\Delta E_Z/\sqrt{3}$.

\subsubsection{Gate noise}

Both single- and two-qubit gate noise are modelled via random variations of the physical parameters of the gate. Namely, as silicon spin qubits are dominated by $1/f$ charge noise~\cite{kepa_2023, elsayed_low_2024, yoneda_noise_correlation_2023, jnane_ab_initio_2024, shehata_2023, yonedaQuantumdotSpinQubit2018, culcer_2009, rojas_spatial_correlations_2023}, we assume that the $g$-factor at every location is randomised at the start of the experiment, but does not change over time. This is valid as long as the time fluctuations are slow compared to the frequency at which the physical components are recalibrated. Therefore, depending on where a qubit is located at a given moment of the experiment --- its position varies when it is shuttled --- its $g$-factor may fluctuate in an uncalibrated manner, which is a source of errors.

For a given perfect gate $U_\text{target}$ implemented as $U(g)$, the fidelity as a function of $g$ reads
\begin{equation} \label{eq:fid_formula}
    \mathcal{F}(g) = \frac{r+|\text{Tr}(U^\dagger_\text{target}U(g))|^2}{r(r+1)}
\end{equation}
where $r=2$ or $4$ for single- or two-qubit gates respectively. We then extract the average fidelity over the distribution $p(g)$ of the $g$-factor,
\begin{equation}
    \overline{\mathcal{F}} = \int \mathcal{F}(g)p(g)~ \mathrm{d}g,
\end{equation}
assuming that the $g$-factor fluctuations follow a normal distribution
\begin{equation}
    p(g) = \frac{1}{\sqrt{2\pi}\sigma_g} \mathrm{e}^{-(g-g_0)^2/2\sigma_g^2}
\end{equation}
where $g_0$ is the expected $g$-factor at the qubit's location, obtained by prior calibration, and $\sigma_g$ its standard deviation. In the QEC simulations we set the average fidelity $\overline{\mathcal{F}}$, and determine the value of $\sigma_g$ that reproduces it. Once this has been evaluated, we sample a set of $g$-factors for each gate location and fix it for the rest of the simulations. This accounts for the fact that the gate properties are defined by their surrounding electromagnetic environment, and assumes that they are static over time. Ultimately, this means that each gate location has a fixed fidelity $\mathcal{F}$, and that only the average $\overline{\mathcal{F}}$ over all gates was initially given. This is the essential difference with the depolarising model of \cref{sec:noise_models}, where a single rate $p$ applies uniformly to every location. Additionally, given the specific structure of each gate's matrix, the weights of the Pauli errors are not uniform.

Note that for CZ gates, we write the noise as a function of the $g$-factor difference $g_{12}$ between the two dots it is acting on. One may also wonder why the $J$ coupling is not assumed to impact the CZ gate noise. While this parameter does fluctuate with charge noise, its influence is only second-order when the detuning between the dots is null~\cite{Reed2016reduced}, which is the working point we are assuming.

For the Hadamard and the CZ respectively, we numerically evaluate the dependence of $\overline{\mathcal{F}}$ on $\sigma_g$ and $\sigma_{g_{12}}$ via Monte-Carlo simulations, and find
\begin{align}
    1-\overline{\mathcal{F}}_\text{H} &= 1.26 \times 10^9 \,(\sigma_g/g)^2, \\
    1-\overline{\mathcal{F}}_\text{CZ} &= 2.14 \,(\sigma_{g_{12}}/g_{12})^2.
\end{align}
The very large prefactor in $\overline{\mathcal{F}}_\text{H}$ arises from the large ratio $B_0/B_1$ of the DC to AC magnetic fields.

It is important to note that the noise arising from this modelling is inherently coherent, which may have detrimental effects on the error correction power of our codes compared to less realistic stochastic noise~\cite{Bravyi_2018}. We however always twirl the noise~\cite{Cai_2019}, without which QEC simulations would not be possible.

Additionally, it is worth mentioning that we neglect any crosstalk and assume that gates can be applied to every qubit locally. In reality, driven single-qubit gates may have to be performed using a global pulse, as it may be extremely challenging to localise a driving magnetic pulse at the scale of an electron ($100$~nm). We leave this to further work.

\subsubsection{Other noise processes}

\paragraph{Shuttling} As in the main text, we model shuttling as pure dephasing noise, parameterised by the shuttling error per micrometre $p_\text{sh}$.

\paragraph{Idling} Idling noise takes the same form as \cref{eq:idle}, where $t$ is the time of the operation that is being waited on~\cite{burkard_2023, yonedaQuantumdotSpinQubit2018}. Data qubits are Hahn-echoed exactly as in \cref{sec:noise_models}.

\paragraph{Initialisation} Qubits are initialised in the $\ket{0}$ state, but may be flipped with a probability $p_\text{init}$.

\paragraph{Readout} Every measurement outcome is classically flipped with a probability $p_\text{meas}$.

To allow a direct comparison with \cref{fig:ler_vs_p_depol}, we sweep all of these quantities together through the single parameter $p$, each operation's average infidelity being set to $p$, and treat $p_\text{sh}$ as an independent parameter as before.

\subsection{Simulations with spin-specific noise}

Using the noise model above, we perform Monte-Carlo simulations aiming at estimating the logical performance of the mitten codes under our noise model and on our architecture. The general workflow is the following: for a given code and noise level, we first sample the local physical parameters at each gate location ($g$-factors). These stay fixed for the entire duration of the experiment, which is consistent with a $1/f$ noise assumption, dominated by low frequencies. We then run between 1,000 and 1,000,000 Monte-Carlo samples (depending on the target logical error rate), where errors are randomly applied to data and ancilla qubits after each operation according to their local $g$-factor, for $d$ rounds of stabiliser measurements. For every run, the information from these measurements is fed to the decoder to compute a correction and determine if an $X$ or $Z$ logical error occurred. We then deduce a logical error rate $P_L$ \textit{i.e.} the probability that either an $X$ or $Z$ logical failure occurs.

\begin{figure}
    \centering
    \includegraphics[width=0.5\linewidth]{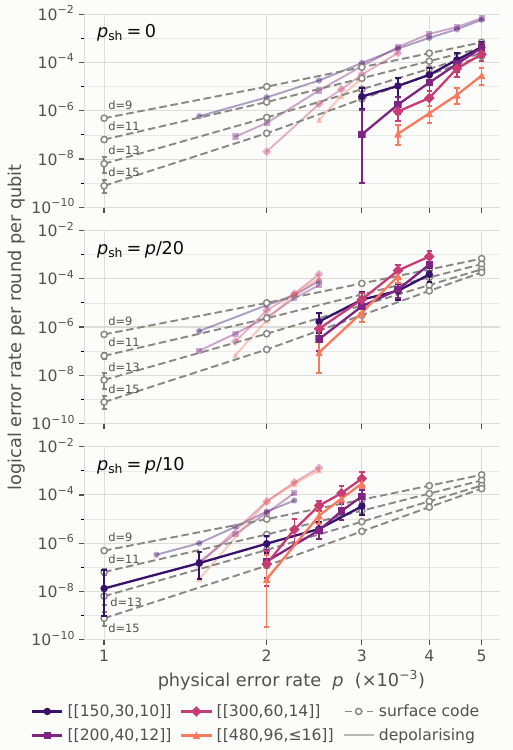}
    \caption{\textbf{Logical error rate simulations under a silicon-spin-qubit-specific noise model.} The figure is the analogous of \cref{fig:ler_vs_p_depol} which was obtained under more generic depolarising noise (and that we repeat here with the faded coloured lines). The logical error rate per round per logical qubit is plotted against the physical error rate $p$, for the first four codes of \cref{tab:target_codes} compiled onto the NeST architecture and simulated for $d$ rounds of syndrome extraction. The three panels differ only in the shuttling dephasing rate, from $p_\mathrm{sh}=0$ (top) to $p_\mathrm{sh}=p/10$ (bottom). Dashed grey lines are surface codes of distance $9$ to $15$ under the same noise model but with no shuttling error, which is conservative in their favour. The vertical dash-dotted line in each panel marks the threshold; it stays above $p=0.1\%$ throughout.}
    \label{fig:ler_vs_p_tailored}
\end{figure}

The results are shown in \cref{fig:ler_vs_p_tailored}, with the depolarising curves of \cref{fig:ler_vs_p_depol} repeated in faded colours for comparison. At equal average physical infidelity, every code performs better under the spin-specific model than under depolarising noise, and the threshold sits correspondingly higher. The reason is that the tailored noise is far more structured. Depolarising noise spreads its weight evenly over all Pauli types and uniformly over all locations, which is the least informative distribution for the decoder. The physical model does neither. Some Pauli patterns may have higher weights than others (\textit{e.g.} the exchange-based CZ has no single-spin-flip error); and it is spatially heterogeneous, since each gate location draws a $g$-factor once and keeps it, so a given location is persistently better or worse than average rather than fluctuating around it. Both features are exploitable by the decoder. The bias concentrates the likely error patterns on a smaller set of configurations, and the static heterogeneity means the prior passed to belief propagation can be made location-dependent, so that the decoder knows in advance which qubits to distrust. Neither piece of information exists in the depolarising model, justifying why we viewed it as conservative in the main text.

Two caveats temper this. First, the coherent character of the underlying noise is only benign because we assume randomised compiling. Second, the decoder must be given location-dependent priors to reach such advantage --- this requires constant monitoring of the device components.

\section{Construction of the NeST-compatible surgery gadget} \label{app:surgery}

\cref{sec:logic} builds a thickened NeST-compatible gadget in two steps: first adding edge ancillas then cycle checks, such that the parity check matrices of the deformed code satisfy the defining guidelines of NeST (see the start of \cref{sec:architecture}). This appendix details both steps for the $\llbracket 150,30,10\rrbracket$ code, then gives the resulting schedule and the checks behind \cref{tab:gadgets}.

We use the presentation of the code released with the gadgets of Ref.~\cite{bhardwaj2026highrateqldpcprocessors}, which differs from that of \cref{tab:target_codes_full} by group translations of the base-matrix entries, and measure the same operator $\bar X_e$ as their seed gadget. Its support has weight $18$, with $17$ qubits in $D_1$ and one in $D_2$, and meets $26$ checks of $Z_0$. The skeleton therefore has $18$ vertex checks, $17$ in $V_1$ and one in $V_2$, and $26$ edge ancillas in $E_{\text{skel}}$.

\subsection{First step: edge ancillas}

The first thickening step chooses a coupling $\alpha=(\alpha_1,\alpha_2)$ and the sites of $E'$ that receive an edge ancilla. $\Gamma^\top$ must take the form of \cref{eq:Gamma_nest_condition}, with as few edge ancillas as possible, and under three requirements:
\begin{itemize}
    \item[$(i)$] the gadget graph has a Cheeger constant $h\geq1$, so that the merged code keeps the distance of the original one;
    \item[$(ii)$] every row of $\Gamma$ has even weight, \textit{i.e.} every edge ancilla meets an even number of vertex checks, so that the product of the vertex checks is $\bar X_e$;
    \item[$(iii)$] the columns of $\Gamma$ have low weight: we require every vertex check, which acts on one data qubit and on a column of $\Gamma$, to have weight at most $10$, the weight of the deformed $Z_0$ checks, so that the gadget does not raise the maximal check weight of the merged code.
\end{itemize}

We write $S_1,S_2\in\mathbb{F}_2[G]$ for the support of $\bar X_e$ in $D_1$ and $D_2$. In \cref{eq:Gamma_nest_condition}, the edge ancilla at site $x$ of $E'$ interacts with the vertex checks of the qubits $x\alpha_1\cap S_1$ and $x\alpha_2\cap S_2$: vertex checks only exist on the support, so that each block $R(\alpha_i)$ is restricted to the rows $S_i$, and to the columns of the sites that receive an ancilla. By $(ii)$, a site may only receive an ancilla if this number of vertex checks is even, and is only useful if it is nonzero: we call such sites usable. A coupling $\alpha$ thus offers a set of usable sites, of which only some receive an ancilla, the others being left empty. The nonempty sites of the skeleton are all usable since $\bar X_e$ commutes with $Z_0$.

Requirement $(i)$ is a covering problem. For a set $T$ of vertex checks, let $\partial T$ be the set of edge ancillas meeting $T$ an odd number of times, which by $(ii)$ is also $\partial(V\setminus T)$. The Cheeger constant is
\begin{equation*}
    h=\min_{\emptyset\neq T\subsetneq V}\frac{|\partial T|}{\min(|T|,|V|-|T|)},
\end{equation*}
so that $h\geq1$ if and only if $|\partial T|\geq\min(|T|,|V|-|T|)$ for every $T$. We compute it exactly by enumerating the $2^{18}$ subsets $T$: the skeleton has $h=1/9$. Each added edge ancilla increases $|\partial T|$ by one for every $T$ it meets an odd number of times, so that reaching $h\geq1$ amounts to covering the deficit $\min(|T|,|V|-|T|)-|\partial T|$ of every $T$, which we solve by integer programming. Without the NeST constraint, $8$ added edges are needed, as in Ref.~\cite{bhardwaj2026highrateqldpcprocessors}, whose $8$ edges however involve $5$ distinct couplings. Imposing a single coupling, we enumerated the $961$ couplings of weight two and three with usable sites, up to left translation, and solved for each the integer program placing the fewest ancillas at usable sites under $(i)$ and $(iii)$. Two couplings of weight two and $107$ of weight three succeed, the best with $8$ edge ancillas, the unconstrained minimum. We retain $\alpha_1=\{e,rs^3\}$, $\alpha_2=\emptyset$, which places $8$ ordinary edges at $8$ of its $11$ usable sites and needs the fewest gate layers.

\subsection{Second step: cycle checks}

The second step chooses the cycle checks and their coupling $\gamma$, such that $\Lambda$ takes the form of \cref{eq:Lambda_nest_condition}, with as few gate layers $|\gamma|$ as possible, under three requirements:
\begin{itemize}
    \item[$(iv)$] $\Lambda\Gamma=0$, \textit{i.e.} every cycle check meets every vertex check an even number of times, so that the merged checks commute;
    \item[$(v)$] the rows of $\Lambda$ span the cycle space of the gadget graph, of dimension $|E|-|V|+1=17$, so that the gadget adds no logical qubit;
    \item[$(vi)$] the rows of $\Lambda$ have low weight: we require every cycle check to have weight at most $10$, so that the gadget does not raise the maximal check weight of the merged code.
\end{itemize}

Cycle checks interact with edge ancillas only. We label both using the larger group $\tilde G=C_{2n_1}\rtimes C_{n_2}=\langle\rho,\sigma\rangle$, with $\sigma\rho\sigma^{-1}=\rho^{-1}$. It contains $G$ as the index-two subgroup $\langle\rho^2,\sigma\rangle$, with $r=\rho^2$ and $s=\sigma$. In \cref{eq:Lambda_nest_condition}, each cycle check sits at its own site $y\in\tilde G$ of a single belt of $2|G|$ sites, and interacts with the edge ancillas at the sites $yk$, $k\in\gamma$: each element of $\gamma$ is one displacement of that belt, \textit{i.e.} one gate layer. Unlike for $\Gamma$, $\Lambda$ is not merely a restriction of $R(\gamma^*)$ to the occupied sites: within a layer, a cycle check may land next to an edge ancilla outside its cycle, and must then hold fire.

By $(iv)$, every cycle check is a cycle of the gadget graph. We therefore draw candidate checks from its short cycles and their translations, each of which may be placed at any free site of the belt. Checks are added greedily, by the number of independent cycles gained per new displacement, until $(v)$ holds, and displacements are then pruned one at a time while the others suffice. Over randomised restarts, this gives $17$ cycle checks, a basis of the cycle space, of weight at most $6$ and with $|\gamma|=12$, an upper bound on the optimum.

\subsection{Merged round}

All edge ancillas sit in the unit cell of $D_1$, which also holds the $17$ vertex checks of $V_1$. There, a merged round runs the memory layers of $X_0$ and $Z_0$, one layer for the deformation of $Z_0$, $9$ vertex-check layers and the $12$ cycle-check layers. $V_1$ takes $6$ of the vertex-check layers, one onto its data qubits and five onto edge ancillas ($|b_0^*|+|\alpha_1|$), and $V_2$ the remaining $3$ ($|b_1^*|$), after acting on its data qubit in its own cell. Every layer is one rigid displacement of one train, and we checked that the CZ gates applied reproduce exactly the nonzero entries of the merged check matrices.

\subsection{Verification}

Our gadget and, for reference, that of Ref.~\cite{bhardwaj2026highrateqldpcprocessors} pass the following checks, which underlie \cref{tab:gadgets}.
\begin{itemize}
    \item \textit{Merged code.} The merged checks commute; $V$ is the only nontrivial set of vertex checks meeting every edge ancilla evenly, so that $\bar X_e$ and nothing else is measured; the cycle checks span the cycle space; and the merged code encodes $k-1=29$ logical qubits.
    \item \textit{Distance.} Integer programming gives $d_X=d_Z=10$ for the original code and for every merged code, each proven optimal within seconds, as guaranteed by $h\geq1$.
    \item \textit{Protocol.} In a noiseless Stim~\cite{gidney2021stim} simulation of two memory rounds, merging, three merged rounds, splitting and two memory rounds, every detector is deterministic, the product of the vertex-check outcomes equals $\bar X_e$, all other logical operators are preserved and $\bar Z_e$ is randomised. This holds for every single-logical flip in both bases and after rewiring to each of the $30$ operators $\bar X_g$.
\end{itemize}

\end{document}